\documentclass[letterpaper]{article}
\pdfoutput=1
\usepackage{jheppub}
\usepackage{amsmath}
\usepackage{graphicx}% Include figure files
\usepackage[config=altsf]{subfig}
\usepackage{feynmf}
\usepackage{dsfont}
\usepackage{array}
\usepackage{slashed}
\usepackage{afterpage}
\usepackage{appendix}
\usepackage{multirow}
\usepackage{wrapfig}
\usepackage[usenames,dvipsnames]{xcolor}
\usepackage[utf8]{inputenc}

\usepackage{dcolumn}
\usepackage{slashed}

\newcommand{\be}{\begin{equation}}
\newcommand{\ee}{\end{equation}}
\newcommand{\bea}{\begin{eqnarray}}
\newcommand{\eea}{\end{eqnarray}}
\newcommand{\lsim}{\!\mathrel{\hbox{\rlap{\lower.55ex \hbox{$\sim$}} \kern-.34em \raise.4ex \hbox{$<$}}}}
\newcommand{\gsim}{\!\mathrel{\hbox{\rlap{\lower.55ex \hbox{$\sim$}} \kern-.34em \raise.4ex \hbox{$>$}}}}

\title{Broadening Dark Matter Searches at the LHC: Mono-X versus Darkonium Channels}
\author[a]{Anirudh Krovi,}
\author[a,b,c]{Ian Low,}
\author[a,d]{and Yue Zhang\,}
\affiliation[a]{Department of Physics and Astronomy, Northwestern University, Evanston, IL 60208, USA}
\affiliation[b]{High Energy Physics Division, Argonne National Laboratory, Argonne, IL 60439, USA}
\affiliation[c]{Theoretical Physics Department, CERN, 1211 Geneva 23, Switzerland}
\affiliation[d]{Theoretical Physics Department, Fermilab, Batavia, IL 60510, USA}

\abstract{Current searches for dark matter at the LHC focus on mono-X signatures: the production of dark matter in association with a Standard Model (SM) particle. The simplest benchmark introduces a massive spin-1 mediator, the $Z^\prime$ boson, between the  dark matter $\chi$ and the SM. Limits derived from mono-X channels are most effective when the mediator can decay into two on-shell dark matter particles:  $M_{Z'}\gtrsim 2M_\chi$. We broaden the experimental reach  into the complementary region, where the $Z^\prime$  mediator is much lighter than the dark matter. In this scenario the $Z^\prime$ mediates an effective long-range force between the dark matter, thereby facilitating the formation of  darkonium bound states, as is common in many dark sector models. The darkonium becomes active when $M_{\chi}>M_{Z'}/\alpha_{\rm eff}$, where $\alpha_{\rm eff}$ is the effective fine-structure constant in the dark sector. Moreover, the darkonium could decay back into SM quarks, without producing missing transverse momentum in the detector. Considering multijet final states, we reinterpret existing  searches to constrain the simple $Z^\prime$ benchmark beyond the region probed by mono-X searches. Assuming a baryonic $Z^\prime$ mediator and a Dirac dark matter, direct detection bounds  can be loosened by giving a small Majorana mass to the dark matter. We also consider the interplay between mono-X and darkonium channels at future high energy colliders, which is at the frontier of probing the model parameter space.}

\preprint{CERN-TH-2018-169, FERMILAB-PUB-18-326-T}

\begin{document}

\maketitle

%%%%%%%%%%%%%%%%%%%%%%%%%%%%%%%%%%%%%%%%%%%%%%%%%%%%%%%%%%%%%%%%%%%%
\section{Introduction}
%%%%%%%%%%%%%%%%%%%%%%%%%%%%%%%%%%%%%%%%%%%%%%%%%%%%%%%%%%%%%%%%%%%%

One important mission of the Large Hadron Collider (LHC) and future high energy colliders is to 
probe the nature of dark matter. If the dark matter particle has a coupling 
to the standard model sector, it could be produced at the LHC, usually in
pairs if the dark matter is stabilized by a (possibly new) symmetry. The dark matter particles are expected to escape 
the detector like neutrinos. They can lead to events with large missing 
transverse momenta, if another visible object ({\it e.g.}, an energetic jet) 
is produced at the same time. The monojet process has been widely studied 
at the Tevatron, LHC and future colliders~\cite{Bai:2010hh, Goodman:2010ku, 
An:2012va}. The same idea has been extend to other standard model particles 
being produced in together with dark matter leading to the so-called mono-X searches~\cite{Abercrombie:2015wmb}.

Going beyond mono-X, another important aspect of dark matter at colliders is
the production of dark bound states. Bound states made of dark 
matter and its anti-particle (darkonium) exist generically in dark sector models 
with a dark force carrier whose  coupling to the dark matter is strong enough.
They are the analog of the positronium or heavy quarkonium states in the real world, 
which have played an instrumental role in our understanding of the SM. It is conceivable that similar phenomena would occur in a dark sector containing the dark matter~\cite{Strassler:2006im, MarchRussell:2008tu, Braaten:2013tza, Laha:2013gva, Wise:2014jva, Wise:2014ola, Laha:2015yoa, An:2016gad, An:2016kie, Cirelli:2016rnw, Braaten:2017gpq, Mitridate:2017izz, Biondini:2017ufr, Baldes:2017gzw, Baldes:2017gzu, Geller:2018biy, Biondini:2018pwp, Biondini:2018xor}.
The signatures of darkonium have been studied at both lepton and 
hadron colliders in several models~\cite{Shepherd:2009sa, An:2015pva, Tsai:2015ugz, Bi:2016gca}.
The bound state formation channel is also an ideal place for probing 
the self-interactions of dark matter in the laboratories~\cite{An:2015pva}.

In this work, we investigate the complementarity between the mono-X and and darkonium
channels in the LHC search for dark matter. Our study is based on a simple renormalizable 
model where a Dirac fermionic dark matter $\chi$ is charged under the gauged baryon number symmetry.
The new  $Z'$ boson mediates the interaction between dark matter and quarks.
This simple model is widely adopted  as the benchmark  for  LHC monojet analyses~\cite{Aaboud:2017phn, Sirunyan:2017hci}. 
So far, experimental limits have been derived in the region of parameter space with $M_{Z'}\gtrsim 2M_\chi$, where the $Z^\prime$ can decay into two on-shell dark matter particles. Outside of this region the production rate of a pair of dark matter particles through the off-shell $Z^\prime$ is too small and the mono-X searches become ineffective. It is possible to directly search for the production of $Z^\prime$ which subsequently decays back into the SM quarks, resulting in multijet final states~\cite{Sirunyan:2017nvi, ATLAS:2016xiv, Sirunyan:2016iap, Aaboud:2017yvp, DijetArgonneTalk}. However, the resonance search in the multijet final states quickly loses its constraining power for a $Z^\prime$ at or below the weak scale, due to the overwhelming QCD background. Therefore, there is presently no experimental search that is sensitive to the $Z^\prime$ benchmark when the $Z^\prime$ is light.

In this work we would like to point out that, in the commonly adopted benchmark for mono-X searches, the $Z'$ boson could mediate a long-range dark force  between dark matter particles, when its mass is light  and coupling to dark matter strong. Then the $\chi\bar\chi$ darkonium bound states could exist in nature and be produced at a high energy collider.  Once produced, the $\chi\bar\chi$ inside the darkonium will eventually find each other and annihilate, causing the latter to be unstable and decay back to SM quarks. The novelty here is that, although the dark matter particle is produced at the collider, there is  are  missing energy/momentum in the final state! In this case, the darkonium would appear as a resonance in multijet final states and its production can be constrained in these searches. In turn, an experimental limit on the production rate of darkonium can be translated into constraints on the mass and couplings of the dark force carrier: the $Z^\prime$. In the end we find the darkonium signals are most active when the  $Z'$-quark coupling is weak and  the  $Z'$-dark-matter coupling is strong. When darkonium exists, it offers a new  handle to explore the nature of dark matter at colliders, and can be highly complementary to the mono-X channel as well as the direct searches for the $Z'$ boson.

This paper is organized as the following. In section~\ref{sec:2}, we describe the simple benchmark model  and discuss the necessary condition for the darkonium bound states to exist, which includes requiring the $Z'$ to be  lighter than the dark matter, precisely the region where the mono-X search in ineffective. We give a brief summary on the existing searches for a light baryonic $Z'$ boson. In section~\ref{LHC}, we calculate the darkonium production cross section and the possible decay channels. We explore the feasibility of using the di-jet channel to search for the darkonium states appearing as new resonances. We derive the existing LHC limit as well as the projections  at the future high-energy high-luminosity LHC, as well as a possible 100 TeV $pp$ colliders. These results are compared with the reach of the monojet channel. We highlight the complementarity of mono-X versus darkonium searches, both of which are needed to effectively cover each other's blind spot. In section~\ref{DMdetections}, we discuss the implications from other areas of dark matter searches,  including direct and indirect detections,  as well as the its production mechanism in the early universe.  We identify the parameter space where high-energy colliders are at the frontier of searching for dark matter in this model.
Then we conclude in section~\ref{sec:4}.

While this paper was being prepared, a related work~\cite{Elor:2018xku} appeared which explored dark matter bound state signals in several non-minimal dark sectors with quite sizable dark couplings. However, the simple benchmark model discussed in this work was not covered.

%%%%%%%%%%%%%%%%%%%%%%%%%%%%%%%%%%%%%%%%%%%%%%%%%%%%%%%%%%%%%%%%%%%%
\section{The Benchmark Model}\label{sec:2}
%%%%%%%%%%%%%%%%%%%%%%%%%%%%%%%%%%%%%%%%%%%%%%%%%%%%%%%%%%%%%%%%%%%%

In mono-X searches the commonly adopted simplified model includes a massive spin-1 boson, the $Z^\prime$, mediating the production of the dark matter particle, which is assumed to be a vector-like pair of fermions $(\chi,\bar{\chi})$. The leading low-energy effective Lagrangian takes the form
\begin{equation}\label{LEFT}
\mathcal{L}_{\rm EFT} = \mathcal{L}_{\rm SM} + g_q \bar q \slashed{Z'} q - \frac{1}{4} Z'_{\mu\nu}Z'^{\mu\nu} + \frac{1}{2} M_{Z'}^2 Z'_\mu Z'^\mu + \bar \chi \left(i\slashed\partial + \left(g_\chi+ g_\chi' \gamma_5\right) \slashed{Z'} - M_\chi \right) \chi \ .
\end{equation}
The $Z'$ is assumed to have a universal coupling $g_q$ to SM quarks and, to be general, we allow for both the vector and axial-vector current couplings with the dark matter. The axial coupling then implies the $Z'$ current is anomalous, which can be remedied by postulating spectator fermions to restore the gauge invariance associated with the $Z'$. We further assume these spectator fermions to be much heavier than the weak scale. Since we will not consider loop-induced processes involving the $Z^\prime$ boson in this work, the anomalous $Z^\prime$ current (or equivalently the anomaly-cancelling spectator fermions) plays no role in our study~\cite{Fox:2018ldq}.

A concrete example of a $Z^\prime$ boson is to gauge  the baryon number symmetry $U(1)_B$ in the SM, which is anomalous with respect
to the electroweak gauge groups. The ultraviolet complete models of gauged baryon number have been 
discussed in~\cite{Carone:1994aa, Carone:1995pu, FileviezPerez:2010gw, Duerr:2013dza, Duerr:2013lka, Perez:2014qfa, Duerr:2014wra, Ohmer:2015lxa}. In Eq.~(\ref{LEFT}) we have also  extended the minimal gauged baryon number model by introducing an additional dark matter field $\chi$ that is charged under
the $U(1)_B$. The presence of the axial coupling $g_\chi'$ implies that $\chi_L$ and $\chi_R$ must carry different charges under the $U(1)_B$.
As a result the dark matter mass $M_\chi$ is not $U(1)_B$ invariant and must be generated via the Yukawa coupling of $\chi$ to the vacuum expectation value (vev) of a $U(1)_B$ Higgs field. The same vev also contributes to the mass of the $Z'$. 
In the appendix~\ref{SCPV}, we present a simple model for this.
Requiring the Yukawa coupling to satisfy the unitarity constraint results in an upper bound on $g_\chi'$ (see Eq.~(\ref{upperlimitg'}) and also~\cite{Kahlhoefer:2015bea}),
\begin{eqnarray}\label{eq:unitarity}
g_\chi' \lesssim \frac{M_{Z'}}{M_\chi}  \ .
\end{eqnarray}
Therefore, if $M_{Z'}\ll M_\chi$, which is the region of interest in this work, $g_\chi'$ need to be small.

For a dark matter interacting with the quark through a $Z'$ mediator, constraints from dark matter direct detection are quite stringent for a light $Z'$~\cite{ArkaniHamed:2008qn, Pospelov:2007mp, Feng:2010zp, Zhang:2015era}. Such constraints could be relaxed by introducing, in addition to the Dirac mass term for $\chi$,  a Majorana mass in Eq.~(\ref{LEFT})~\cite{TuckerSmith:2001hy, TuckerSmith:2004jv, DeSimone:2010tf, Zhang:2016dck},
\begin{equation}\label{LEFT2}
\frac{\delta}{2} \overline{\chi^c} \chi + {\rm h.c.} \ ,
\end{equation}
We assume $\delta$ is small enough compared to the Dirac mass $M_\chi$ so that our discussions on collider phenomenology in section~\ref{LHC} remain
valid at the zeroth order in the small $\delta$ expansion, which allows us to treat $\chi$ as a Dirac fermion in collider studies. On the other hand, $\delta$ must be large enough to evade the direct detection constraints. A quantitative estimate of $\delta$ satisfying both considerations will be presented in section~\ref{DMdetections}. A non-vanishing $\delta$ will have  implications in cosmology and indirect detection of dark matter, which will also be explored in section~\ref{DMdetections}.

%%%%%%%%%%%%%%%%%%%%%%%%%%%%%%%%%%%%%%%%%%%%%%%%%%%%%%%%%%%%%%%%%%%%
\subsection{The Formation of Darkonium}\label{introBS}
%%%%%%%%%%%%%%%%%%%%%%%%%%%%%%%%%%%%%%%%%%%%%%%%%%%%%%%%%%%%%%%%%%%%

One important aspect of dark matter we want to explore  is bound state physics. The $Z'$ exchange yields a  Yukawa potential between $\chi$ and $\bar \chi$. With a light enough $Z'$ and large enough couplings $g_\chi, g_\chi'$, bound states made of $\chi$ and $\bar\chi$ could form. 
Because of the fermionic nature of $\chi$, there are two darkonium ground states, one with total spin $S=0$ and the other with $S=1$, which we denote by  $\eta_D$ and $\Upsilon_D$, respectively.  We will focus on ground states in this work.

The vector coupling of the $Z'$ with the dark matter yields an attractive Yukawa potential while the sign of the potential from the axial coupling depends on the total spin \cite{Bellazzini:2013foa}. We can define the effective fine-structure constant of the $Z'$-mediated long range interaction between the dark matter particles as
\begin{equation}
\alpha_{\rm eff}(S) = \alpha_\chi + \frac{4}{3}\left(S(S+1) - \frac{3}{2}\right) \alpha_\chi' 
\end{equation}
where $\alpha_\chi = g_\chi^2/(4\pi)$, $\alpha_\chi' = g_\chi'^2/(4\pi)$.
The potential is  attractive for $S=1$ and repulsive for $S=0$. The $Z'$ boson plays two roles in this model. It is not only the mediator between the dark matter and the SM, but also the dark force carrier responsible for self-interactions of the dark matter.

At the LHC, the darkonium can be created via an off-shell $Z'$ boson, much like the production of $J/\Psi$ particle through an off-shell photon in QCD. Therefore, the spin-1 darkonium $\Upsilon_D$ can be singly produced on resonance, 
while the spin-0 darkonium $\eta_D$ has to be produced in association with another $Z'$.\footnote{This is different from fixed energy colliders where the  $\eta_D$ and $\Upsilon_D$ channels are comparably important~\cite{An:2015pva}.}
In what follows we will focus on the spin-1 darkonium $\Upsilon_D$, in which case 
\begin{equation}\label{aeff}
\alpha_{\rm eff} = \alpha_\chi + \frac{2}{3} \alpha_\chi' \ .
\end{equation}
Then the condition for the ground state to exist is~\cite{RogersGraboskeHarwood},
\begin{equation}\label{eq:exist}
\frac{\alpha_{\rm eff} M_\chi}{M_{Z'}} > 1.68 \simeq \frac{\pi^2}{6} \ .
\end{equation}
The mass of $\Upsilon_D$ is given by $2M_\chi$ minus the ground state binding energy, $BE$. In the Coulomb limit ($M_{Z'}\to0$),
$BE = \alpha_{\rm eff}^2 \mu/2$, where $\mu$ is the reduced mass of the system 
\begin{equation}\label{eq:mu1}
\mu=\frac{1}{2} M_\chi  \ .
\end{equation}
For general nonzero $M_{Z'}$, the binding energy can be solved numerically~\cite{RogersGraboskeHarwood}. A useful analytic approximation 
can be obtained using the Hulth\'en potential to mimic the Yukawa potential~\cite{LamVarshni}. 
In this case, 
\begin{equation}
\label{eq:BEdef}
BE \simeq \frac{\alpha_{\rm eff}^2 \mu}{2} \left( 1 - \frac{\pi^2}{12} M_{Z'} a_0 \right)^2 \ ,
\end{equation}
where $a_0 \equiv 1/({\alpha_{\rm eff} \mu})$. 
One could also derive the bound state wavefuction at the origin, which is~\cite{LamVarshni},
\begin{equation}\label{wave}
\Psi(0) \simeq \sqrt{\frac{1 - \left({\pi^2}M_{Z'} a_0/12 \right)^2 }{\pi a_0^{3}}} \ .
\end{equation}

The single production of $\Upsilon_D$ at the LHC could be described, effectively, by a kinetic mixing with  the $Z'$ boson, which takes the form~\cite{An:2015pva}
\begin{equation}\label{eq:KineticMixing}
\mathcal{L}_{\Upsilon_D-Z'\,\rm mixing} = \frac{\kappa}{2} Z'_{\mu\nu} \Upsilon_D^{\mu\nu} \ , 
\end{equation}
where
\begin{equation}\label{kappa}
\kappa = \sqrt{\frac{2\pi \alpha_\chi}{M_\chi^3}}\Psi(0)  \ .
\end{equation}
Through this kinetic mixing $\Upsilon_D$ could couple to SM quarks.  Any non-zero axial current coupling will introduce further kinematic mixings of both the $Z'$ and $\Upsilon_D$ (known as the $1^{--}$ ground state) with the $1^{++}$ state, an excited bound state made of $\chi\bar\chi$. However, these mixings are suppressed compared to Eq.~(\ref{eq:KineticMixing}) by additional powers of $\alpha_{\rm eff}$. We will therefore truncate the spectrum and only consider the ground state for the rest of this paper. 

%%%%%%%%%%%%%%%%%%%%%%%%%%%%%%%%%%%%%%%%%%%%%%%%%%%%%%%%%%%%%%%%%%%%
\subsection{Current Constraints on the $Z'$ Mediator}
%%%%%%%%%%%%%%%%%%%%%%%%%%%%%%%%%%%%%%%%%%%%%%%%%%%%%%%%%%%%%%%%%%%%

Experimentally, a vector boson $Z'$ that couples to SM quarks could be produced at hadron colliders such as the Tevatron and LHC. If the $Z'$ is lighter than
twice of the dark matter mass, it can only decay back to a SM quark and antiquark. Existing dijet resonance searches cover the $Z'$ mass 
window from 50\,GeV to multiple TeV scales. Below we list several limits from the recent analysis on the $Z'$-quark-qntiquark coupling
from dijet searches (see also~\cite{DijetArgonneTalk}).
\begin{table}[h]
\centering\begin{tabular}{c|c|c|c}
\hline
CMS 13\,TeV~\cite{Sirunyan:2017nvi} & $35.9\,{\rm fb}^{-1}$ & $50\,{\rm GeV} \lesssim M_{Z'} \lesssim 300\,{\rm GeV}$ & $g_q\lesssim 0.06-0.2$ \\ 
ATLAS 13\,TeV~\cite{ATLAS:2016xiv} & $3.4\,{\rm fb}^{-1}$ &  $450\,{\rm GeV} \lesssim M_{Z'} \lesssim 950\,{\rm GeV}$ & $g_q\lesssim 0.06-0.14$ \\ 
CMS 13\,TeV~\cite{Sirunyan:2016iap} & $12.9\,{\rm fb}^{-1}$ & $600\,{\rm GeV} \lesssim M_{Z'} \lesssim 3500\,{\rm GeV}$ & $g_q\lesssim 0.07-0.44$\\
ATLAS 13\,TeV~\cite{Aaboud:2017yvp} & $37.0\,{\rm fb}^{-1}$ & $1.5\,{\rm TeV} \lesssim M_{Z'} \lesssim 3.5\,{\rm TeV}$  & $g_q\lesssim 0.07-0.27$  \\ 
\hline
\end{tabular}
%\caption{}\label{LimitTable}
\end{table}\\
These limits directly apply to the  $Z'$ in our model when it predominantly decays into $q\bar q$~\footnote{If the $Z'$ is heavier than twice of the dark matter mass, there are mono-X constraints which will be reviewed briefly in the next section.}. 
The future running of LHC and the dijet searches could further improve the coverage of $Z'$ mass from 50\,GeV up to a few TeV,
leaving the region of light $Z'$ below 50\,GeV as a blind splot.

Through the quark loops, the $Z'$ boson mixes with the SM $Z$-boson. As a result, when the masses of the two are close enough, there are 
useful limits from the hadronic $Z$-boson width measurement at the LEP~\cite{Graesser:2011vj, Dobrescu:2014fca}. 
For even lighter $Z'$, below a few GeV, there are also constraints on its mixing with the heavy quarkonium
states like the $\Upsilon$ and $J/\Psi$, as well as the rare decay of meson states into the $Z'$. For a recent study, see~\cite{Tulin:2014tya}.
It is also worth noting that for very light bayonic $Z'$, the heavy anomalon fields can have strong non-decoupling effects on 
flavor-changing neutral current processes~\cite{Dror:2017ehi, Dror:2017nsg}.
However, our study here will mainly focus on the region $M_{Z'}>10$ GeV, thus these non-decoupling effects can be evaded.

In the next section, we will show that the same search results could be reinterpreted as constraints on the production of the darkonium $\Upsilon_D$, leading to new limits on the dark matter simplified model that are complementary to the mono-X searches.

%%%%%%%%%%%%%%%%%%%%%%%%%%%%%%%%%%%%%%%%%%%%%%%%%%%%%%%%%%%%%%%%%%%%
\section{Darkonium Versus Mono-X}\label{LHC}
%%%%%%%%%%%%%%%%%%%%%%%%%%%%%%%%%%%%%%%%%%%%%%%%%%%%%%%%%%%%%%%%%%%%

\begin{figure}[t]
\centerline{\includegraphics[width=0.8\textwidth]{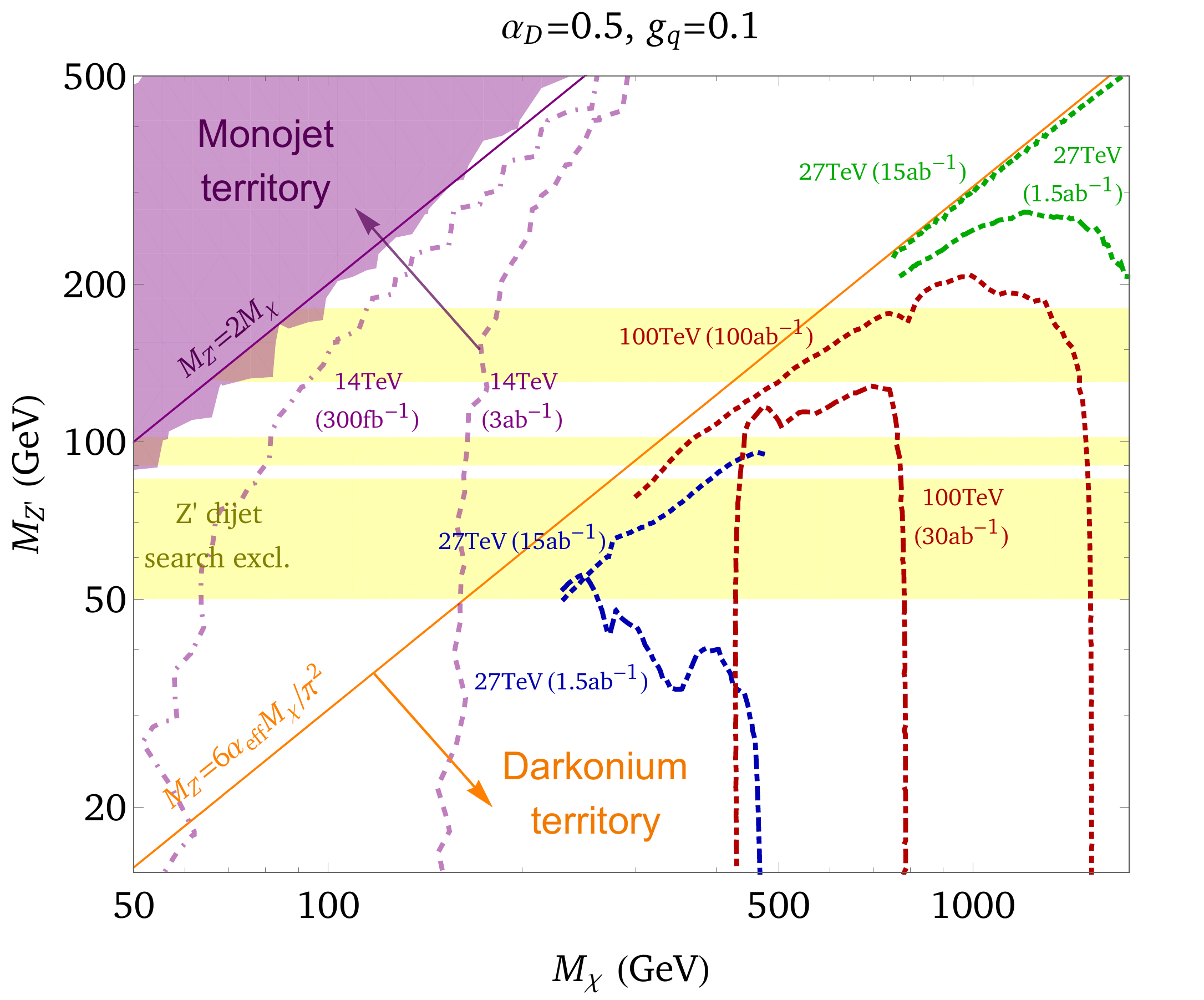}}
\caption{Colorful curves show the future high-energy $pp$ collider constraints on the model where a fermionic dark matter $\chi$ interacts with SM quarks via a $Z'$ boson. 
For the couplings defined in Eq.~(\ref{LEFT}), we set $g_q=0.1$, $\alpha_\chi=0.5$ and $g_\chi'=M_{Z'}/M_\chi$.
The regions to the left of the cyan curves (monojet search) and below the blue, red, green curves (darkonium resonance search) could be covered.
The horizontal yellow bands are excluded by the existing dijet search for $Z'$. In the future, this search will cover all the region above $M_{Z'}>50\,$GeV but below the orange line.
}\label{AllConstraints}
\end{figure}

In this section, we will explore the interplay between darkonium and mono-X channels in searches for dark matter whose interaction is mediated by a 
$U(1)_B$ baryonic vector boson. They turn out to be highly complementary to each other in probing the model parameter space. 
Moreover, the dijet $Z'$ search at LHC seems to have a blind spot for light $Z'$ below 50\,GeV.
As explained above, a $Z'$ is light enough could facilitate the existence of darkonium bound states. Search for the formation of such new states at the LHC 
could in turn constrain the light $Z'$ as a dark force and help covering the above blind spot.
These important features are summarized in Fig.~\ref{AllConstraints}. We will go through the details of this plot for the rest of this section. 
Generically, the mono-X searches are most sensitive to the region $M_{Z'}\gtrsim 2M_\chi$, while the darkonium searches
mainly probe the region $M_{Z'} \lesssim \alpha_{\rm eff}M_{\chi}$.

Hereafter, we will choose the follow benchmark values for the model parameters,
\begin{equation}\label{BenchmarkParameters}
\alpha_\chi = 0.5, \ \ \ g_q=0.1, \ \ \ g_\chi'=\frac{M_{Z'}}{M_\chi} \ .
\end{equation}
Before moving on, we comment on how our results change when the benchmark values of $\alpha_\chi$ and $g_q$ vary from the 
choice in Eq.~(\ref{BenchmarkParameters}).
First, from Eq.~(\ref{eq:DY}) (see below), the dark matter bound state production cross section
is proportional to $\alpha_\chi^4$. The monojet cross section (dominate by ISR jet radiation) is proportional to $\alpha_\chi$.
Their limits will get substantially weaker for smaller values of $\alpha_\chi$, especially for the bound state channel.
We also restrict ourselves to the region with $\alpha_\chi<1$ so that the $\chi$ particles in the bound state are still non-relativistic and we could reliably do 
perturbative calculations in the small $\alpha_\chi$ expansion. 
Second, we choose a relatively smaller value of $g_q$ than commonly used in the previous monojet analysis (where $g_q\gtrsim0.2$ is used).
This is mainly driven by the increasingly stronger bound from the $Z'$ search in the dijet channel. 
With a coupling $g_q\gtrsim0.2$, most the region in Fig.~\ref{AllConstraints} with $M_{Z'}>50\,$GeV is already excluded by the current LHC data.

We will scan the rest of parameter space and present our results in the $M_{Z'}$ versus $M_{\chi}$ plane.  We find that, with the current LHC data (13\,TeV, $\sim36\,{\rm fb}^{-1}$), 
we are not yet able to derive a competitive limit in the parameter space of interest. However, future experiments such as the upcoming high luminosity running of LHC at 14\,TeV (expected luminosity up to $\sim3\,{\rm ab}^{-1}$), the high-energy high-luminosity LHC running at 27\,TeV (expected luminosity up to $\sim15\,{\rm ab}^{-1}$~\cite{HLLHCTalk}), 
and a possible  100\,TeV hadron collider~\cite{Hinchliffe:2015qma}, will enable us to derive very useful limit in the parameter space
where the dark matter bound states could be produced. In Fig.~\ref{AllConstraints}, we show the region of parameter space that could be probed by these future experimental programs.

%%%%%%%%%%%%%%%%%%%%%%%%%%%%%%%%%%%%%%%%%%%%%%%%%%%%%%%%%%%%%%%%%%%%
\subsection{Mono-X Searches}
%%%%%%%%%%%%%%%%%%%%%%%%%%%%%%%%%%%%%%%%%%%%%%%%%%%%%%%%%%%%%%%%%%%%

\begin{figure}[t]
\centerline{\includegraphics[width=5cm]{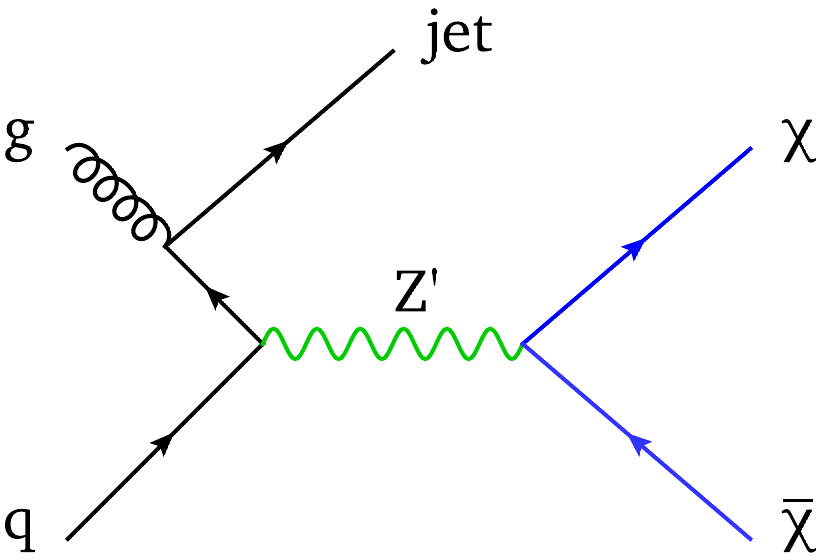}\hspace{1cm}\includegraphics[width=6.4cm]{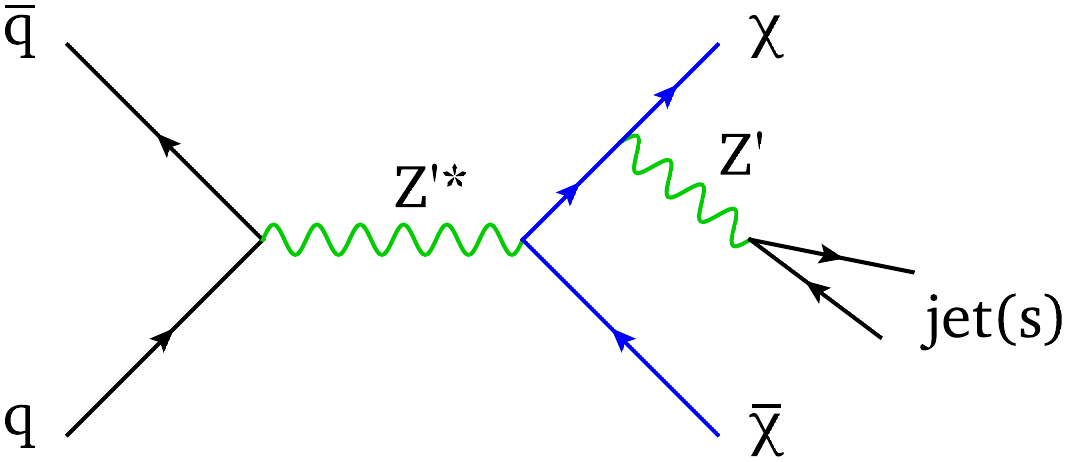}}
\caption{Feynman diagram for a monojet event in $\chi\bar\chi$ production at LHC, due to initial state (left) and final (right) state radiations.}\label{fig:monojet}
\end{figure}

The simple model in Eq.~(\ref{LEFT}) has served as the benchmark model for many mono-X searches for dark matter at the LHC, where the dark matter particles $\chi$ and $\bar\chi$ are produced in together with a SM particles. In particular, the ``monojet'' final states are characterized with very large transverse missing energy (MET) plus one or more jets. A representative Feynman diagram is shown in Fig.~\ref{fig:monojet} (left). The $qg$ initiated process is dominant because of the large gluon parton distribution function (PDF) at small $x$. In the parameter space where $M_{Z'} > 2 M_\chi$,
the $Z'$ boson could be produced on-shell in association with one or more jets, $q g \to q + Z'$, followed by the decay $Z'\to \chi\bar\chi$ resulting in MET. 

Recent monojet analyses of the model by ATLAS and CMS collaborations can be found in Refs.~\cite{Aaboud:2017phn, Sirunyan:2017hci}. With $g_q=0.1$ and $\alpha_D=0.5$, constraints derived from rescaling the results in Refs.~\cite{Aaboud:2017phn, Sirunyan:2017hci} do not yet place useful limits on the  parameter space shown  in Fig.~\ref{AllConstraints}. However, according to our estimates, this will change with the future running of the LHC at higher luminosities (300\,${\rm fb}^{-1}$ and 3\,${\rm ab}^{-1}$). The reaches are shown by the cyan dot-dashed and dotted curves in Fig.~\ref{AllConstraints}. The region to the left of these curves could be covered.  With a large enough luminosity, the monojet constraint extends slightly into the region where $M_{Z'}<2M_\chi$, where the monojet production is a $2\to3$ process, $qg\to q\chi\bar\chi$, with the $Z'$ being off-shell. 
Nevertheless, the cross section decreases rapidly with increasing dark matter mass,  because the radiated jet needs to
have a large transverse momentum, of order a few hundred GeV, to satisfy the experimental trigger. This feature limits the ability of using monojet channel to probe the parameter space deep in the $M_{Z'}<2M_\chi$ region.

Instead of initial state jet radiation, one may also consider final state radiation of the $Z'$ boson, $q\bar q \to \chi\bar\chi Z'$. In the $M_{Z'}<2M_\chi$  region, the $Z'$ can only decay back to $q\bar q$, which appear as two jets. For a sufficiently light and boosted $Z'$, the two jets will be collimated with each other and may appear as a single jet in the detector. In this case one could apply the monojet analysis to this channel. However, the final state radiation process must be initiated by $q\bar q$ initial states, see Fig.~\ref{fig:monojet} (right), and the cross section is suppressed by the anti-quark PDF over the gluon PDF compared to initial state radiation case.  We include this channel in our analyses and find the modification to the total monojet cross section to be small (less than 10\%). It is possible to study this channel further by  exploring the possible jet substructure~\cite{Bai:2015nfa}, as well as displaced vertex~\cite{Izaguirre:2015zva} signatures.

%%%%%%%%%%%%%%%%%%%%%%%%%%%%%%%%%%%%%%%%%%%%%%%%%%%%%%%%%%%%%%%%%%%%
\subsection{Darkonium Searches}\label{sec:DMBSLHC}
%%%%%%%%%%%%%%%%%%%%%%%%%%%%%%%%%%%%%%%%%%%%%%%%%%%%%%%%%%%%%%%%%%%%

The limitation of mono-X searches outside the $M_{Z^\prime} < 2 M_{\chi}$ region strongly motivates us to consider additional possible dark matter production channels at the LHC, in particular,  bound states of $\chi, \bar\chi$. These states are unstable and will decay promptly back (the decay rates are given by Eq.~(\ref{eq:UpsilonDecay})) to SM quarks, appearing as dijet (or multi-jet) resonances,  which lead to very different collider signatures from monojet.  
We want to emphasize again that the darkonium search here is different from direct searches of $Z^\prime$ as dijet resonances discussed in Section 2.2.
Here the $Z'$ plays the role of a dark force for the darkonium to exist. The darkonium production cross section is proportional to its wavefunction at the origin thus depends on the $Z'$ mass and couplings. Constraining the formation of such darkonium state allows us to indirectly constrain the dark force.
We discuss these in detail in this subsection.

%%%%%%%%%%%%%%%%%%%%%%%%%%%%%%%%%%%%%%%%%%%%%%%%%%%%%%%%%%%%%%%%%%%%
\subsubsection{$\Upsilon_D$ production}
%%%%%%%%%%%%%%%%%%%%%%%%%%%%%%%%%%%%%%%%%%%%%%%%%%%%%%%%%%%%%%%%%%%%

\begin{figure}[t]
\centerline{
\includegraphics[width=0.8\textwidth]{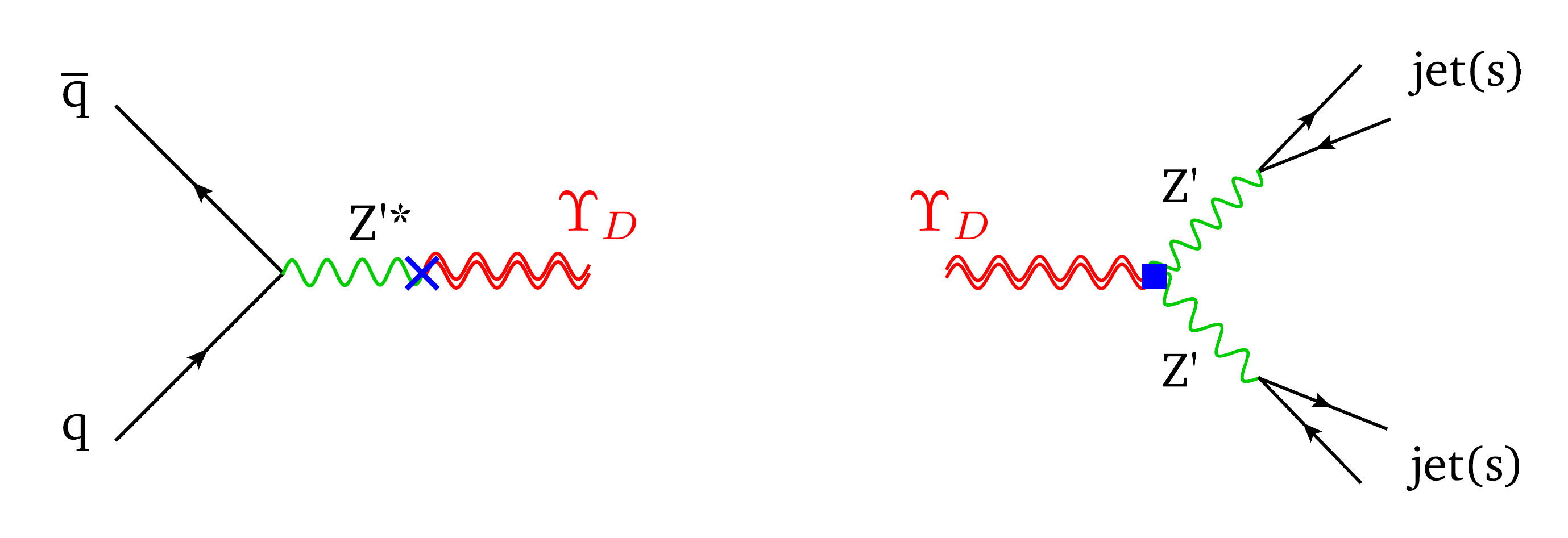}}
\caption{{\sf Left}: Feynman diagram for darkounium state $\Upsilon_D$ production at LHC.
The blue cross represents the $\Upsilon_D$-$Z'$ kinetic mixing given in Eq.~(\ref{eq:KineticMixing}).
{\sf Right}: Feynman diagram for one of the $\Upsilon_D$ decay channels, into two $Z'$ bosons, with the latter cascade decay into jet(s). 
The blue square represents the $\Upsilon_D$-$Z'$-$Z'$ vertex given in Eq.~(\ref{eq:UpsilonZ'Z'}).
}\label{fig:twoZdecay}
\end{figure}

\begin{figure}[t]
\centerline{\includegraphics[width=0.7\textwidth]{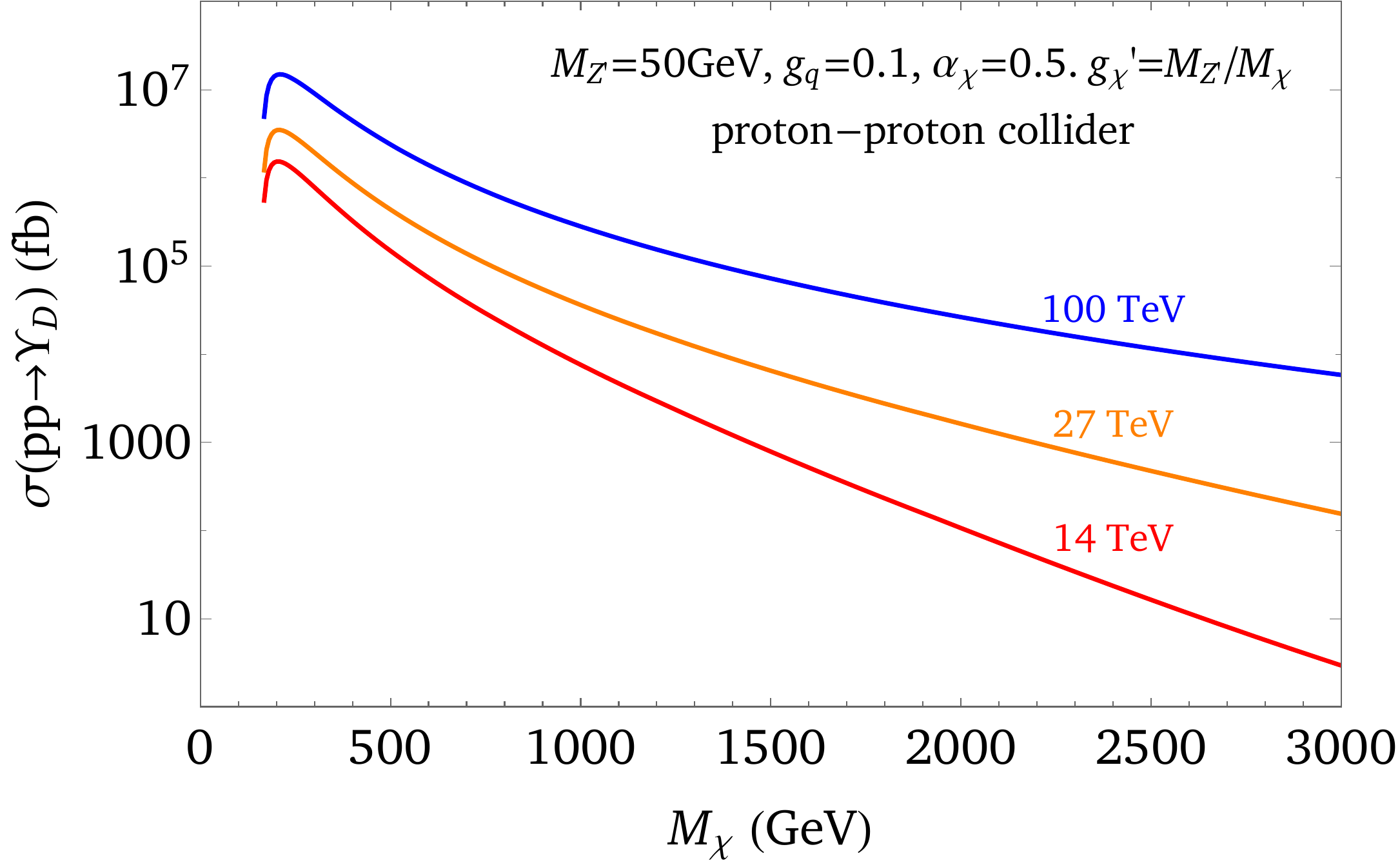}}
\caption{Production cross section of $\Upsilon_D$ at the LHC as a function of the dark matter mass $M_\chi$.}\label{DrellYan}
\end{figure}

As discussed in section~\ref{introBS}, we will focus on the spin-1 darkonium state $\Upsilon_D$ at the LHC. It is mainly produced via $q\bar q$ fusion and the Feynman diagram is shown in Fig.~\ref{fig:twoZdecay} (left). The production cross section at a proton-proton collider takes the form
\begin{equation}\label{eq:DY}
\sigma_{pp\to \Upsilon_D}= \frac{\pi \kappa^2 g_q^2}{s_{\rm CM}}\left(\frac{4M^{2}_{\chi}}{4M^{2}_{\chi}-M^{2}_{Z'}}\right)^{2}  \sum_q \int_{\tau}^1\frac{dx}{x} 
\left[ f_{q/p}(x) f_{\bar q /p}\left(\frac{\tau}{x}  \right) + f_{\bar q/p}(x) f_{ q /p}\left(\frac{\tau}{x}\right)\right]  \ .
\end{equation}
where $\tau=M_{\Upsilon_D}^2/s_{\rm CM}$, $s_{\rm CM}$ is the center-of-mass energy of $pp$ collision, and the parameter $\kappa$ is given by Eq.~(\ref{kappa}).
In Fig.~\ref{DrellYan}, we plot this cross section at various collider energies ($\sqrt{s_{\rm CM}} =14, 27, 100$\,TeV) as a function of the dark matter mass,
$M_\chi$, with $M_{Z'}=50\,$GeV and the other parameter fixed as in Eq.~(\ref{BenchmarkParameters}). 
The choice of $g_\chi'$ value follows from the consideration in Eq.~(\ref{eq:unitarity}).
Here we calculated the cross section using the {\tt NNPFD}~\cite{Ball:2014uwa} with the PDF set {\tt NNPDF30\_lo\_as\_0118\_nf\_6}.
After the production, $\Upsilon_D$ will decay into two (or more) jets as will be discussed in the section~\ref{sec:decay}. 
We will use the dijet resonance search data to set limits and estimate future reach at the LHC and higher energy colliders.
\\

%%%%%%%%%%%%%%%%%%%%%%%%%%%%%%%%%%%%%%%%%%%%%%%%%%%%%%%%%%%%%%%%%%%%
\subsubsection{$\Upsilon_D$ decay}\label{sec:decay}
%%%%%%%%%%%%%%%%%%%%%%%%%%%%%%%%%%%%%%%%%%%%%%%%%%%%%%%%%%%%%%%%%%%%

After production, there are three ways for the darkonium $\Upsilon_D$ to decay: 1) to $q\bar q$ via an off-shell $Z'$; 2) to two $Z'$ bosons; 3) to three $Z'$ bosons. The partial decay rates are
\begin{eqnarray}\label{eq:UpsilonDecay}
\begin{split}
&\Gamma_{\Upsilon_D\to q\bar q} = \frac{N_{f} g_q^2 g_\chi^2}{\pi}\frac{4 M^{2}_{\chi}}{(4M^{2}_{\chi}-M^{2}_{Z'})^{2}} \Psi(0)^{2} \ , \\
&\Gamma_{\Upsilon_D\to 2Z'} = \frac{8 g_\chi^2 g_\chi'^2 (M_\chi^2 - M_{Z'}^2)^{5/2}}{3\pi M_\chi M_{Z'}^2 (2M_\chi^2 - M_{Z'}^2)^2}  \Psi(0)^2 \ , \\
&\Gamma_{\Upsilon_D\to 3Z'} \approx \frac{(\pi^2-9) g_\chi^6}{36\pi^3 M_\chi^2}  \Psi(0)^2  \ ,
\end{split}
\end{eqnarray}
where $N_f$ is the number quark flavors that $\Upsilon_D$ can decay into, and $\Psi(0)$ is given in Eq.~(\ref{wave}). The calculation of non-relativistic bound state decay is reviewed in~\cite{Keung:2017kot}.
The first decay channel is simply the inverse process of the Feynman diagram in Fig.~\ref{fig:twoZdecay} (left).
The second decay channel $\Upsilon_D\to Z'Z'$ is possible only in the presence of nonzero $g_\chi'$ coupling, which violates the charge-conjugation ($C$) parity.
%and it is followed by $Z'$ decaying into $q\bar q$ each. 
The Feynman diagram for this process is shown in Fig.~\ref{fig:twoZdecay} (right).
The effective operator responsible for this decay channel is~\cite{Keung:2008ve}
\begin{equation}\label{eq:UpsilonZ'Z'}
\hat O_{\Upsilon_D\to 2Z'} = \varepsilon_{\mu\nu\alpha\beta} \Upsilon_D^\mu Z'^\nu Z'^{\alpha\beta} \ .
\end{equation}
For the $\Upsilon_D\to 3Z'$ decay rate, we work in the limit that $g_\chi\gg g_\chi'$ and $M_{Z'}\ll M_\chi$. This allows us to derive an analytic expression for the decay rate, in analogy to that of $\Upsilon\to3\gamma$ decay in the SM~\cite{Voloshin:2007dx}. 
Using the value of $g_\chi'$ from Eq.~(\ref{BenchmarkParameters}), we find that $\Gamma_{\Upsilon_D\to 2Z'} \gg \Gamma_{\Upsilon_D\to 3Z'}$, {\it i.e.}, the three-$Z'$ decay is always subdominant.
%Therefore, we will not take it into account in the following analysis.

%%%%%%%%%%%%%%%%%%%%%%%%%%%%%%%%%%%%%%%%%%%%%%%%%%%%%%%%%%%%%%%%%%%%
\subsubsection{Dijet resonance search for $\Upsilon_D$}
%%%%%%%%%%%%%%%%%%%%%%%%%%%%%%%%%%%%%%%%%%%%%%%%%%%%%%%%%%%%%%%%%%%%

With the above production and decay channels, we are now ready to quantify the experimental constraints for  $\Upsilon_D$ by recasting dijet resonance searches.
To date, the ATLAS and CMS collaborations have published several results on the dijet resonance search~\cite{Sirunyan:2017nvi, Sirunyan:2016iap, ATLAS:2016xiv, Aaboud:2017yvp}, which covers the resonance mass from $\sim 50\,$GeV to multiple TeV scales.
These searches assume the heavy resonance to have 100\% decay branching ratio into $q\bar q$.

\begin{figure}[t]
\centerline{\includegraphics[width=1\textwidth]{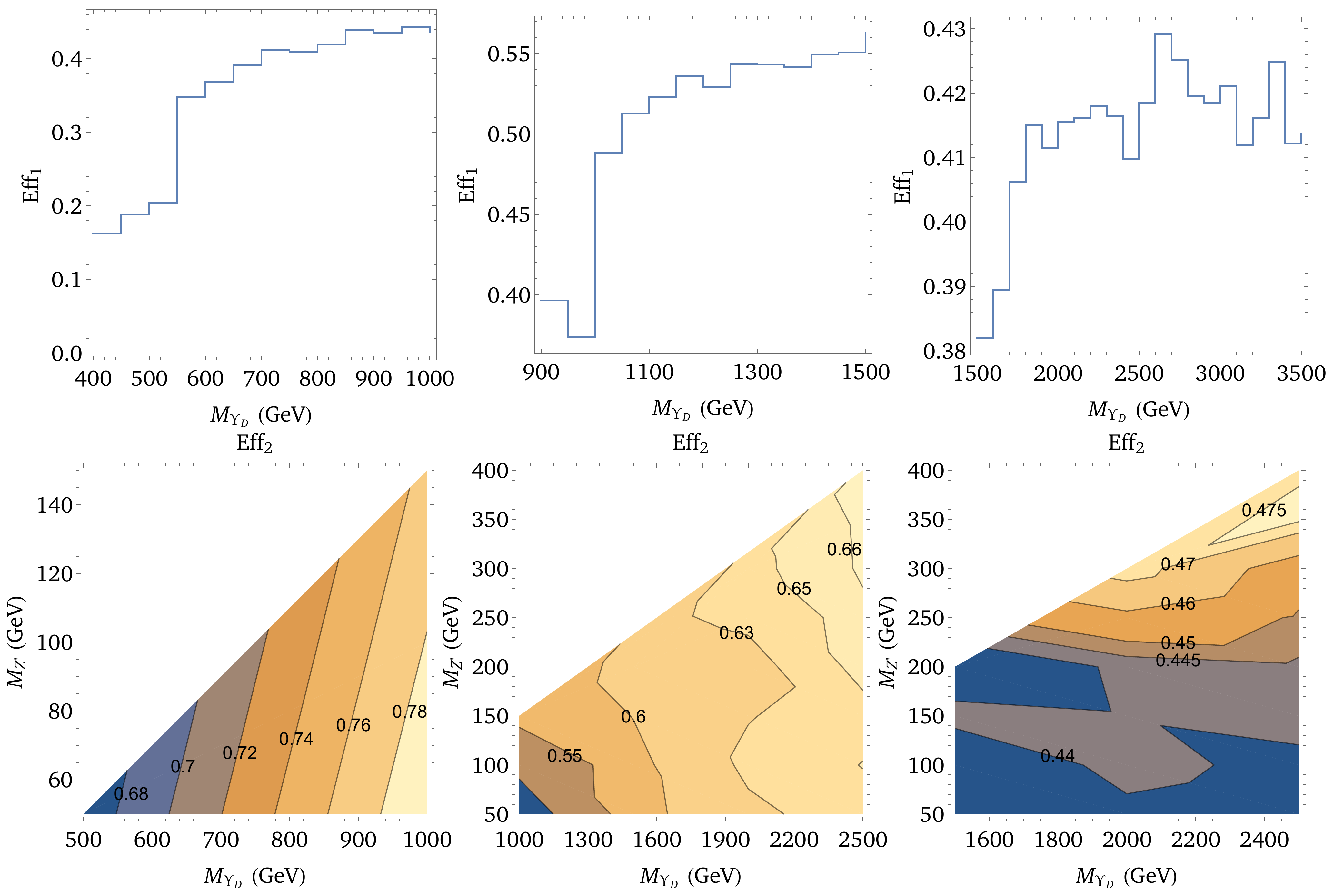}}
\caption{The efficiency factors for the $\Upsilon_D\to q\bar q$ (${\rm Eff}_1$, first tow) and $\Upsilon_D\to 2Z'\to 2(q\bar q)$ (${\rm Eff}_2$, second row) channels for passing the event selection cuts in three mass windows of dijet resonance searches for new vector boson.
}\label{fig:eff}
\end{figure}

However, in our model, $\Upsilon_D$, as the heavy resonance, has more than one decay channels. In order to properly interpret the LHC limits from dijet resonance searches, we need to simulate the selection efficiency of each possible decay channel of $\Upsilon_D$ in Eq.~(\ref{eq:UpsilonDecay}). To this end we first create a {\tt FeynRules}~\cite{Alloul:2013bka} model containing both the $Z^\prime$ boson and  the spin-1 darkonium $\Upsilon_D$. In the model file we  include the kinetic mixing in   Eq.~(\ref{eq:KineticMixing}) responsible for the production of  $\Upsilon_D$, as well as the effective coupling in Eq.~(\ref{eq:UpsilonZ'Z'}) that mediates the  $\Upsilon_D$ decay. Then we use {\tt MadGraph\,5}~\cite{Alwall:2011uj} to generate the $\Upsilon_D$ production and decay to jets at $pp$ colliders, 
and run {\tt PYTHIA\,8}~\cite{Sjostrand:2007gs} and {\tt DELPHES\,3}~\cite{deFavereau:2013fsa} for hadronization and detector simulations.
We follow the dijet event selection cuts described in~\cite{Sirunyan:2017nvi, Sirunyan:2016iap, ATLAS:2016xiv, Aaboud:2017yvp} to derive the efficiency factor, ${\rm Eff}_i$, for each $\Upsilon_D$ decay channel. 
In Fig.~\ref{fig:eff}, we show the efficiency factors for the $\Upsilon_D\to q\bar q$ and $\Upsilon_D\to 2Z'\to 2(q\bar q)$ channels to pass the event selection cuts in each mass window, which are called ${\rm Eff}_1$ and ${\rm Eff}_2$, respectively. We simulate the production of $\Upsilon_D$ at the LHC and take into account of its boost on event-by-event basis. A lighter $\Upsilon_D$ is typically born with a higher boost, thus when it decays the opening angle of final states tends to be smaller, leading to a lower efficiency factor. Such an effect is shown in the upper left panel of Fig.~\ref{fig:eff}.

It is worthwhile remarking on the dijet efficiency factor for $\Upsilon_D\to 2Z'\to 2(q\bar q)$ decay, which is the following.
Kinematically, when $M_{\Upsilon_D} \gg M_{Z'}$, the $Z'$ bosons from the decay of $\Upsilon_D$ are boosted. For an $\Upsilon_D$ produced at rest, 
the two jets from each $Z'$ have a maximal opening angle
\begin{equation}
(\Delta R_{jj})_{\rm max} = 2 \arctan \frac{M_{Z'}}{M_\chi} \ .
\end{equation}
The formation of darkonium requires $M_{Z'}\leq 6 \alpha_{\rm eff} M_{\chi}/\pi^2 \approx 0.6\, \alpha_{\rm eff}\, M_{\chi}$, which leads to $(\Delta R_{jj})_{\rm max}\simeq0.6$ with the benchmark parameters.
If the two jets are within the cone size of $\theta_{jj}<0.4$, they will be reconstructed as a single jet typically. 
There is an order 1 chance for this to occur. This estimate is confirmed by the plots in the second row of Fig.~\ref{fig:eff}.

We find it convenient to define the effective coupling between $\Upsilon_D$ and SM quarks
\begin{equation}
g_{\Upsilon_D} = g_q \kappa \sqrt{\frac{\sum_i {\rm Br_i} \times {\rm Eff}_i}{{\rm Eff}_1}} \ ,
\end{equation}
where $i$ goes through all the possible $\Upsilon_D$ decay channels labelled in Eq.~(\ref{eq:UpsilonDecay}). The sub-label ''1'' stands for the $\Upsilon_D\to q\bar q$ decay channel. The upper limit on the effective coupling $g_{\Upsilon}$ can be directly read from the existing LHC 
limits on elementary $Z'$-quark-antiquark coupling obtained in~\cite{Sirunyan:2017nvi, Sirunyan:2016iap, ATLAS:2016xiv, Aaboud:2017yvp}, for four mass windows (which is called $g_q$ there). 
Because $g_{\Upsilon_D}$ is a function of the all model parameters in Eq.~(\ref{BenchmarkParameters}),
an upper limit on $g_{\Upsilon_D}$ will translate into a contour in the parameter space in Fig.~\ref{AllConstraints}. 
We find that the current LHC data are not yet able to provide a competitive constraint in the plot.
However, the further running of high energy high luminosity LHC (at 27\,TeV), as well as the possible 100\,TeV collider will do.
To estimate the future reaches, we first scale the number of events with the increasing integrated luminosities,
by a factor 
\begin{equation}
R_{\rm lum} = \mathcal{L}_{\rm future}\left/\mathcal{L}_{\rm now} \right. \ , 
\end{equation}
where $\mathcal{L}_{\rm now}$ are given in~\cite{Sirunyan:2017nvi, Sirunyan:2016iap, ATLAS:2016xiv, Aaboud:2017yvp}.
We then calculate the enhancement factors in the production cross sections for both the signal, 
\begin{equation}
R_{\sqrt{s_{\rm CM}}}^{\rm sig} = \sigma_{\sqrt{s_{\rm CM}}}^{\rm sig}\left/\sigma^{\rm sig}_{\rm 13\,{\rm TeV}} \right. \ ,
\end{equation} 
and the background, 
\begin{equation}
R_{\sqrt{s_{\rm CM}}}^{\rm bkg} = \sigma_{\sqrt{s_{\rm CM}}}^{\rm bkg}\left/\sigma_{\rm 13\,{\rm TeV}}^{\rm bkg} \right. \ , 
\end{equation} 
and consider ${\sqrt{s_{\rm CM}}}=14, 27, 100\,$TeV as the future collider energies. 
The $\Upsilon_D$ production cross section is given by Eq.~(\ref{eq:DY}).
The QCD background cross section for dijet production at parton level goes as, $\sim \hat{s}^{-1}$. 
Note that the dijet search is a bump hunt. In practice, we focus on a narrow dijet invariant mass window $\hat{s} \sim M_{\Upsilon_D}^2$.
As a result, the proton-proton level cross sections are proportional to the following quantities (the parton luminosity defined in~\cite{Quigg:2009gg}), respectively
\begin{eqnarray}
\begin{split}
\sigma_{\sqrt{s_{\rm CM}}}^{\rm sig} &\propto \frac{1}{s_{\rm CM}} \sum_q \int_{\tau}^1\frac{dx}{x}  f_{q/p}(x) f_{\bar q/p}\left(\frac{\tau}{x}  \right)  \ , \\
\sigma_{\sqrt{s_{\rm CM}}}^{\rm bkg} &\propto \frac{1}{s_{\rm CM}} \sum_{q} \int_{\tau}^1\frac{dx}{x} 
\left[ f_{q/p}(x) f_{\bar q/p}\left(\frac{\tau}{x}  \right) + f_{g/p}(x) f_{q/p}\left(\frac{\tau}{x}  \right) +  f_{g/p}(x) f_{\bar q/p}\left(\frac{\tau}{x} \right)\right] \\
& \hspace{0.4cm}+ \frac{1}{2s_{\rm CM}} \int_{\tau}^1\frac{dx}{x} 
\left[ f_{g/p}(x) f_{g/p}\left(\frac{\tau}{x}  \right) + f_{q/p}(x) f_{q/p}\left(\frac{\tau}{x}  \right) + f_{\bar q/p}(x) f_{\bar q/p}\left(\frac{\tau}{x}  \right) \right] \ ,
\end{split}
\end{eqnarray}
where $\tau=M_{\Upsilon_D}^2/s_{\rm CM}$.
We calculate the rescaling factors $R_{\sqrt{s_{\rm CM}}}^{\rm sig}$ and $R_{\sqrt{s_{\rm CM}}}^{\rm bkg}$ using the {\tt NNPDF}.

Therefore, the future upper bound on $g_{\Upsilon_D}$ is expected to get stronger by a factor of 
\begin{equation}
\sqrt{
\sqrt{ R_{\rm lum} }\frac{R_{\sqrt{s_{\rm CM}}}^{\rm sig}}{\sqrt{R_{\sqrt{s_{\rm CM}}}^{\rm bkg}}}
} \ .
\end{equation}
The future collider reaches are shown in Fig.~\ref{AllConstraints} for three of the mass windows (blue, red, green curves, with texts next to them denoting the corresponding future collider energy and luminosity). The regions below these curves could potentially be covered. \\[2pt]

%%%%%%%%%%%%%%%%%%%%%%%%%%%%%%%%%%%%%%%%%%%%%%%%%%%%%%%%%%%%%%%%%%%%
\subsubsection{Impact of the Majorana mass term on collider phenomenology}
%%%%%%%%%%%%%%%%%%%%%%%%%%%%%%%%%%%%%%%%%%%%%%%%%%%%%%%%%%%%%%%%%%%%

\begin{figure}[h]
\centerline{\includegraphics[width=0.5\textwidth]{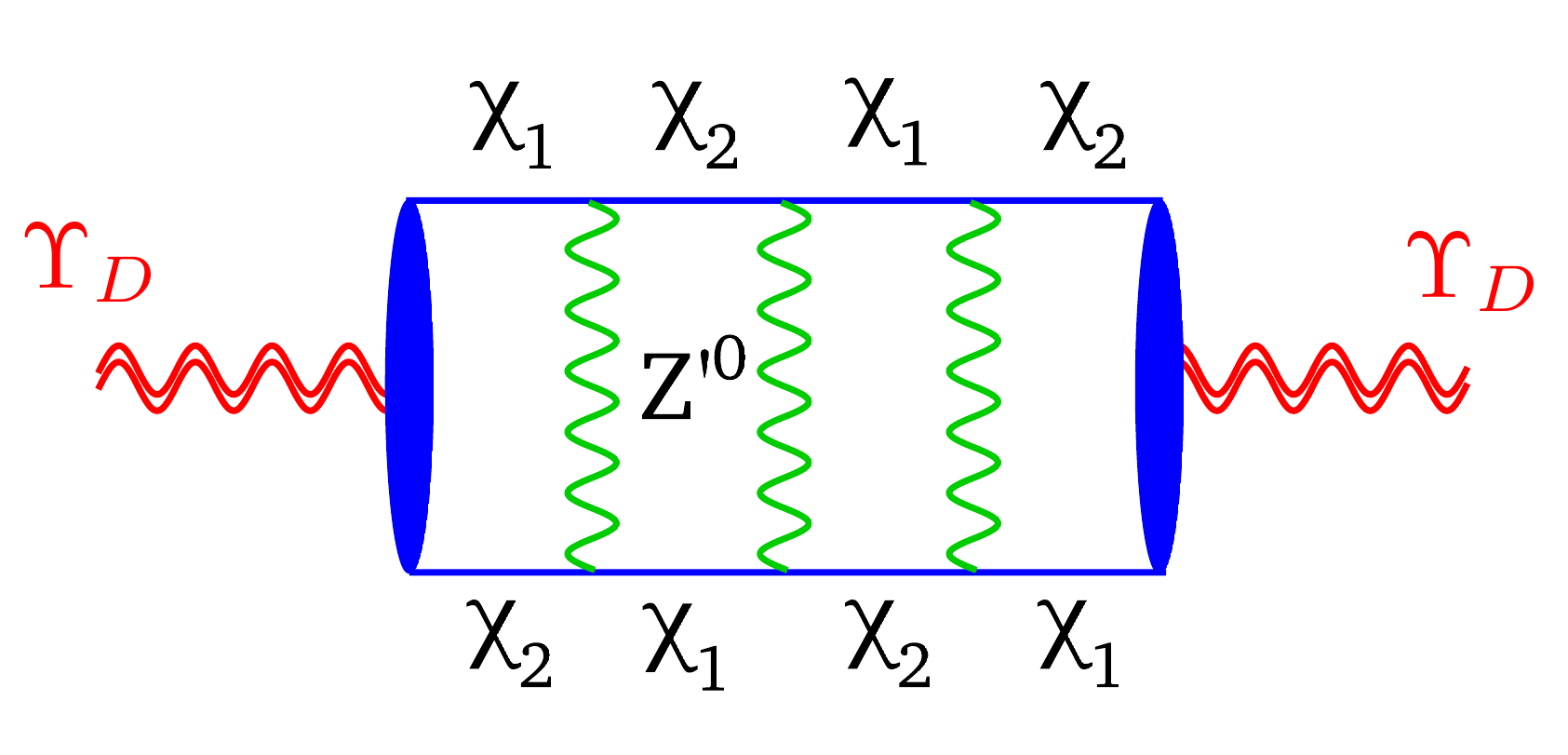}}
\caption{Feynman diagram for a $\chi_1$-$\chi_2$ bound state.
}\label{MajoranaBound}
\end{figure}

So far, our discussions of collider phenomenology are based on the effective Lagrangian Eq.~(\ref{LEFT}) but with the Majorana mass term for $\chi$ defined in Eq.~(\ref{LEFT2})
set to zero. Here we clarify the impact of a nonzero $\delta$ on the dark matter spectrum and the bound state physics LHC.
In the presence of both $M_\chi$ and $\delta$, the mass terms for $\chi$ can be written as
\begin{equation}
- \frac{1}{2} (\bar\chi, \bar\chi^c) \left(\begin{array}{cc}
M_\chi & \delta \\
\delta & M_\chi
\end{array} \right) \left(\begin{array}{c}
\chi \\
\chi^c
\end{array} \right) = - \frac{1}{4} (\bar\chi, \bar\chi^c) 
\left(\begin{array}{cc}
-i & 1 \\
i & 1
\end{array} \right) 
\left(\begin{array}{cc}
M_\chi-\delta & 0 \\
0 & M_\chi+\delta
\end{array} \right) 
\left(\begin{array}{cc}
i & -i \\
1 & 1
\end{array} \right) 
\left(\begin{array}{c}
\chi \\
\chi^c
\end{array} \right)  \ .
\end{equation}
Here we assume $\delta$ is a real parameter.
The two {\it Majorana} fermion mass eigenstates and the corresponding eigenvalues are
\begin{eqnarray}\label{chi12}
\chi_{1} = \frac{i}{\sqrt2}(\chi - \chi^c), \hspace{0.5cm}\chi_{2} = \frac{1}{\sqrt2}(\chi + \chi^c), \hspace{0.5cm} M_{\chi_{1,2}} = M_\chi \mp \delta \ .
\end{eqnarray}
In terms of $\chi_{1,2}$ fields, their interaction terms involving the $Z'$ boson now take the form
\begin{eqnarray}\label{12int}
\mathcal{L}_{Z'\text{-int}} = \frac{1}{2} Z_\mu' (\bar\chi_1, \bar\chi_2) \left(\begin{array}{cc}
g_\chi' \gamma^\mu\gamma_5 & i g_\chi \gamma^\mu \\
-i g_\chi \gamma^\mu & g_\chi' \gamma^\mu\gamma_5
\end{array} \right) 
\left(\begin{array}{c}
\chi_1 \\
\chi_2
\end{array} \right) \ .
\end{eqnarray}
In the parameter space of interest to bound state physics, $M_{Z'}\ll M_{\chi}$, the constraint Eq.~(\ref{eq:unitarity}) indicates that the diagonal axial-current
interactions are suppressed, $g_\chi'\ll g_\chi$. The vector-current interactions are dominant and they must be off-diagonal with respect to $\chi_{1,2}$.
In this case, bound states made of a $\chi_1$ and a $\chi_2$ particle can still form~\cite{Zhang:2016dck}. The long range force due to $Z'$ exchange alternates
the two states along each fermion line (see Fig.~\ref{MajoranaBound}). 

\begin{figure}[t]
\centerline{\includegraphics[width=0.7\textwidth]{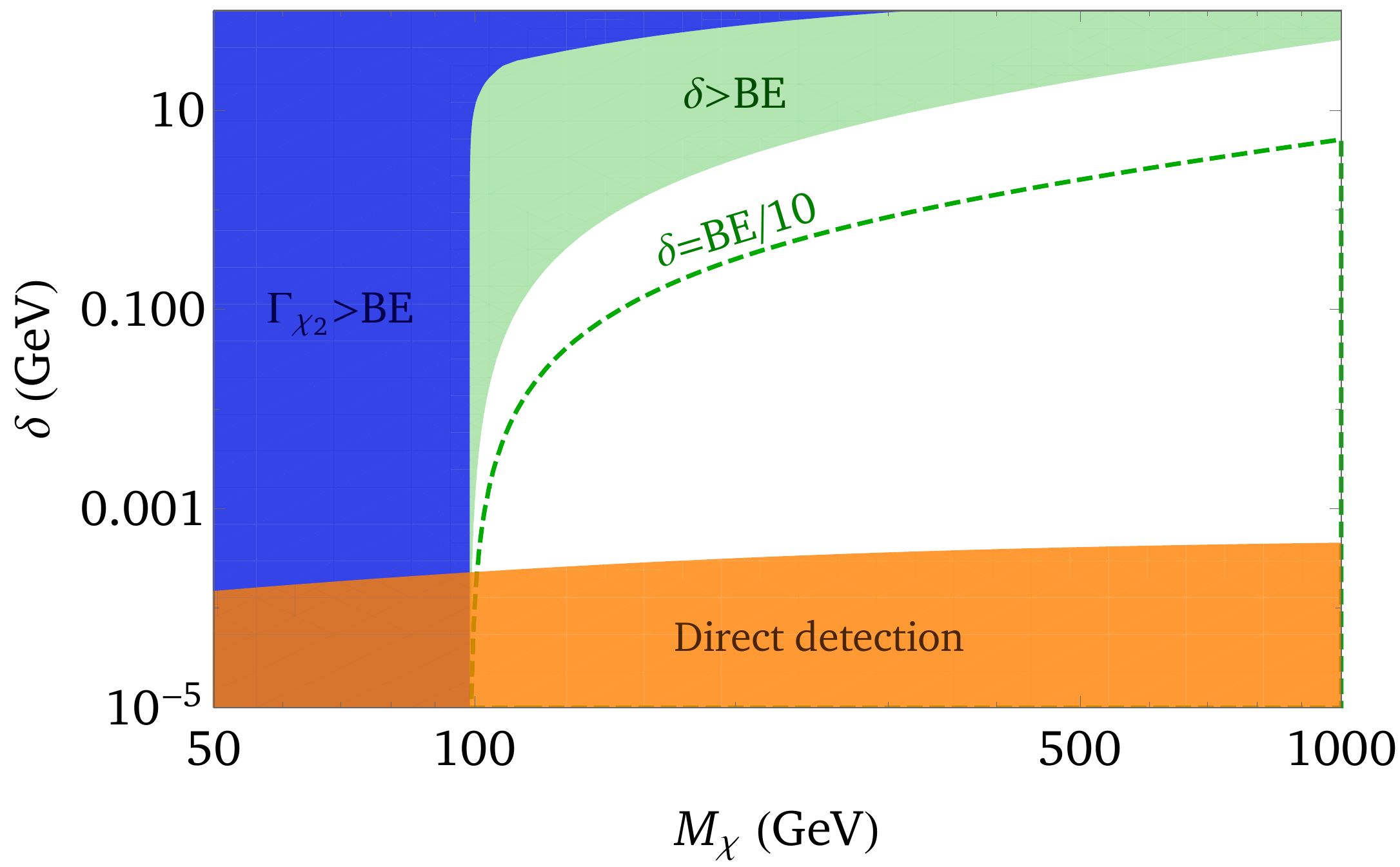}}
\caption{The white region corresponds the parameter space where our discussions on bound state collider physics (see section~\ref{sec:DMBSLHC}) remain valid
in the presence of a nonzero $\delta$, and where the dark matter candidate $\chi_1$ satisfies the direct detection constraints (see section~\ref{DD}).
$M_{Z'}=30\,$GeV here.
}\label{Validity}
\end{figure}

One can still write down a Schr\"odinger equation describing such a bound state,
with the reduced mass now defined as
\begin{equation}
\mu = \frac{M_\chi^2 - \delta^2}{2M_\chi} \ .
\end{equation} 
Using this reduced mass instead of that in Eq.~(\ref{eq:mu1}),
one can repeat the discussions in section~\ref{introBS} to find the spectrum and wavefunctions. 
In the small $\delta/BE$ expansion, where the binding energy $BE$ is defined in Eq.~(\ref{eq:BEdef}), the two results must agree at the leading order.
In Fig.~\ref{Validity} we show in green the region of parameter space with $\delta > BE$. Outside of the green we could have $\delta \ll BE \sim \alpha_\chi^2 M_\chi$. For example, we show in green dashed line in Fig.~\ref{Validity} where $\delta/BE = 1/10$. We expect the main results on bound state collider phenomenology,  derived based on pure-Dirac fermion assumption in the previous subsections, remain unaltered.

This said, in the presence of nonzero $\delta$, the $\chi_2$ particle becomes unstable. 
For $\delta > M_{Z'}$, the following decay could occur, $\chi_2 \to \chi_1 Z'$, whose decay rate is
\begin{equation}
\Gamma_{\chi_2 \to \chi_1 Z'}  = \frac{g_\chi^2\left[ \left( M_{\chi_1}+M_{\chi_2} \right)^2 + 2 M_{Z'}^2 \right] \sqrt{ 
 \left(M_{\chi_1}+M_{\chi_2} \right)^2 - M_{Z'}^2}
}{16\pi M_{Z'}^2 M_{\chi_2}^3} \left(\delta^2 - M_{Z'}^2 \right)^{3/2} \ .
\end{equation}
For $\Lambda_{\rm QCD} \ll \delta < M_{Z'}$, the decay of $\chi_2$ has to occur through off-shell $Z'$, $\chi_2 \to \chi_1 q\bar q$.
In the case $\delta \ll M_\chi, M_{Z'}$, the decay rate takes the approximate form
\begin{equation}
\Gamma_{\chi_2 \to \chi_1 q\bar q} \simeq \frac{N_f g_{q}^{2}g_{\chi}^{2}}{20\pi^{3}M_{Z'}^{4}}\delta^{5} + \mathcal{O}(\delta^6)\ .
\end{equation}
For $\delta\lesssim \Lambda_{\rm QCD}$, the final state $q\bar q$ will turn into meson states.  Isospin singlet vector mesons can directly mix with the baryonic $Z'$ boson.
The decay rate for $\chi_2 \to \chi_1 \omega$ is
\begin{equation}
\Gamma_{\chi_2 \to \chi_1 \omega} = \frac{g_\chi^2 g_q^2 f_\omega^2\left[ \left( M_{\chi_1}+M_{\chi_2} \right)^2 + 2 m_\omega^2 \right] \sqrt{ 
 \left(M_{\chi_1}+M_{\chi_2} \right)^2 - m_\omega^2}}{8\pi M_{Z'}^4 M_{\chi_2}^3} \left(\delta^2 - m_\omega^2 \right)^{3/2} \ ,
\end{equation}
where $f_\omega\simeq70\,$MeV is the decay constant of the $\omega$ meson, $\langle \omega| \bar u \gamma^\mu u + \bar d \gamma^\mu d | 0 \rangle = \sqrt2 f_\omega m_\omega \varepsilon^\mu_\omega$.
For $\delta<m_\omega$, $\chi_2$ could decay into $\chi_1$ plus pions via off-shell $\omega$; and for $\delta<2m_\pi$, $\chi_2$ has to decay into $\chi_1$ plus $e^+e^-$ (or $\mu^+\mu^-$)
through the (loop generated) kinetic mixing between $Z'$ and the photon. 
In practice, we require $\chi_2$ must not decay within the time scale of the bound state formation, which is equivalent to requiring $\Gamma_{\chi_2}$
to be smaller than the bound state binding energy.
For this reason, in Fig.~\ref{Validity}, we also shade out the region with $\Gamma_2 > BE$ in blue color.

%%%%%%%%%%%%%%%%%%%%%%%%%%%%%%%%%%%%%%%%%%%%%%%%%%%%%%%%%%%%%%%%%%%%
\section{Direct, Indirect Detections and Early Universe}\label{DMdetections}
%%%%%%%%%%%%%%%%%%%%%%%%%%%%%%%%%%%%%%%%%%%%%%%%%%%%%%%%%%%%%%%%%%%%

In this section, we discuss the implication of dark matter direct and indirect detection constraints on the model parameter space which was explored in the previous section, using the same set of benchmark parameters given in Eq.~(\ref{BenchmarkParameters}). We also address the possible (thermal) origin of our dark matter relic abundance from the early universe. %Here  we will continue to base  the set of benchmark parameters 

%%%%%%%%%%%%%%%%%%%%%%%%%%%%%%%%%%%%%%%%%%%%%%%%%%%%%%%%%%%%%%%%%%%%
\subsection{Direct detection}\label{DD}
%%%%%%%%%%%%%%%%%%%%%%%%%%%%%%%%%%%%%%%%%%%%%%%%%%%%%%%%%%%%%%%%%%%%

We first consider dark matter direct detection, in the presence of a nonzero $\delta$ parameter. In this case, the dark mater splits into two Majorana 
mass eigenstates, $\chi_1$ and $\chi_2$. Without loss of generality, we assume $\chi_1$ is the lighter one and exist in nature as the dark matter.
$\chi_2$ is an unstable partner state. With the $Z'$-quark-antiquark coupling in Eq.~(\ref{LEFT}) and the $Z'$-$\chi_1$-$\chi_{1,2}$ 
couplings in Eq~(\ref{12int}), there are two types of $\chi_1$-nucleus scattering processes. One is spin-independent and inelastic, $\chi_1 + N \to \chi_2 + N$,
whose cross section is proportional to the product of couplings, $g_\chi^2 g_q^2$. 
The other process is elastic, $\chi_1 + N \to \chi_1 + N$, whose cross section is proportional to $g_\chi'^2 g_q^2$, and depends on the spin of $\chi_1$. Because the SM quarks still couple to $Z'$ coherently via their number density, the spin vector of $\chi_1$ has to be contracted with either its velocity $\vec{v}$, or the three momentum transfer $\vec{q}$. As a result, the cross section is also velocity dependent and suppressed by the halo velocity squared ($v_{\rm halo}\sim10^{-3}c$).
This suppression makes the latter cross section safely small in view of the current direct detection limits.

Next, we will examine the inelastic scattering more carefully. The nucleus-level scattering cross section in the small $\delta$ limit is
\begin{eqnarray}\label{eq:DDxsec}
\sigma^{\rm SI}_{\chi_1 + T \to \chi_2 + T} \simeq  \frac{(3Z)^2 g_q^2 g_\chi^2 \mu_{1T}^2}{\pi M_{Z'}^{4}} \sqrt{1-\frac{2\delta}{\mu_{1T} v_{\rm halo}^2}}\Theta\left(1-\frac{2\delta}{\mu_{1T} v_{\rm halo}^2}\right)\ ,
\end{eqnarray}
where $T$ is the target nucleus, and $\mu_{1T} = M_{\chi_{1}} M_T/(M_{\chi_{1}} + M_T)$. We also assume $M_{Z'}$ is much larger than the momentum transfer of the scattering. The state-of-art dark matter direct detection limits are obtained by the PandaX-II~\cite{Tan:2016zwf}, LUX~\cite{Akerib:2016vxi} and XENON1T~\cite{Aprile:2017iyp} collaborations, where for dark matter mass
of a few hundred GeV, the upper limit on the nucleon-level cross section is $\sigma^{\rm SI}\lesssim 10^{-45}\,{\rm cm}^2$.
The nucleon level scattering cross section can be calculated as
\begin{eqnarray}
\sigma^{\rm SI}_{\chi_1 N} = \frac{\sigma^{\rm SI}_{\chi_1 + T \to \chi_2 + T}}{A^2} \frac{\mu_{1N}^2}{\mu_{1T}^2} \ ,
\end{eqnarray}
where $\mu_{1N} = M_{\chi_{1}} M_N/(M_{\chi_{1}} + M_N)$, $M_N$ is the nucleon mass.
With the benchmark parameters given in Eq.~(\ref{BenchmarkParameters}) and $M_{\chi}=500\,$GeV, $M_{Z'}=50\,$GeV, we find that in the $\delta\to0$ limit,
$\sigma^{\rm SI}_{\chi_1 N} \simeq 10^{-38}\,{\rm cm}^2$, which is much larger than the current upper bounds.~\footnote{This cross section is still too large
given the fact that the relic abundance of $\chi_1$ could be underproduced in a thermal history, where we find $\Omega_{\chi_1}/\Omega_{\rm DM}^{obs}>10^{-5}$ for most of the parameter space (see discussions in the next section  for more details).}
If this was the case, most of the parameter space shown in Fig.~\ref{AllConstraints} would have been ruled out, where we explored the LHC searches for dark matter.

The only way to suppress this cross section and get around the constraint is to turn on $\delta$. The phase space factor in Eq.~(\ref{eq:DDxsec}) implies
a minimal $\chi_1$ velocity for the scattering to occur, $v \geq v_{\rm min} = \sqrt{2\delta/\mu_{1T}}$~\cite{TuckerSmith:2001hy, TuckerSmith:2004jv}.
The usual assumption is that the halo dark matter velocities satisfy the Maxwell-Boltzmann distribution which is peaked around $v_{\rm peak}\simeq270$\,km/s and has a cutoff
at the escape velocity $v_{\rm esc}\simeq 544\,$km/s~\cite{McCabe:2010zh}.
Therefore, if the minimal velocity $v_{\rm min} \gg v_{\rm peak}$, the population of $\chi_1$ that could trigger the scattering process is exponentially suppressed, and if 
$v_{\rm min} > v_{\rm esc}$ the process will be turned off completely. 

We have calculated the lower bound on $\delta$ numerically so that the direct detection limits are satisfied, which is shown by the orange region in Fig.~\ref{Validity}. Approximately, this bound coincides with the kinematic limit,
\begin{eqnarray}\label{eq:MinimalDelta}
\delta \geq \frac{1}{2}\mu_{1T} v^2_{\rm esc} \ .
\end{eqnarray}
The main message from Fig.~\ref{Validity} is that there exist a large window of $\delta$ (the white region) where our collider discussions remain valid and the direct detection constraints are evaded.

%%%%%%%%%%%%%%%%%%%%%%%%%%%%%%%%%%%%%%%%%%%%%%%%%%%%%%%%%%%%%%%%%%%%
\subsection{Thermal relic abundance}
%%%%%%%%%%%%%%%%%%%%%%%%%%%%%%%%%%%%%%%%%%%%%%%%%%%%%%%%%%%%%%%%%%%%

Next, we discuss the dark matter relic abundance in this model. 
We will make the most modest assumption that the dark matter $\chi_1$ and the SM particles were in thermal equilibrium with each other in the early universe.
Its relic abundance is obtained thermally via the freeze out mechanism. 

There are several ways for $\chi_1$ to annihilate in the early universe.
When $M_{\chi_1}>M_{Z'}$, two $\chi_1$ particles can annihilate into two $Z'$ bosons via a $t$- (or $u$-) channel $\chi_2$ exchange.
The annihilation cross section is given by
\begin{equation}\label{eq:Chi1Chi1ToZ'Z'}
(\sigma v)_{\chi_1\chi_1\to Z'Z'} = \frac{
 \left( M_{\chi_1}^2 - M_{Z'}^2 \right)^{3/2}
}{4\pi M_{\chi_1} \left(M_{\chi_1}^2 + M_{\chi_2}^2 - M_{Z'}^2\right)^2} 
\left[ \left(g_\chi^4-6g_\chi^2 g_\chi'^2 + g_\chi'^4\right) + 8  g_\chi^2 g_\chi'^2 \frac{M_{\chi_1}^2}{M_{Z'}^2} \right] \ .
\end{equation}
When $M_{\chi_1}<M_{Z'}$, the above annihilation channel is forbidden, unless one or both of the $Z'$ bosons goes off-shell.
We take into account another channel where two $\chi_1$ particles annihilate into $q\bar q$ via an $s$-channel off-shell $Z'$.
This is only possible via the diagonal $Z'$ coupling in Eq.~(\ref{12int}) which is an axial-current interaction involving two $\chi_1$ particles.
Its cross section is given by, 
\begin{equation}\label{eq:Chi1Chi1Toqqbar}
(\sigma v)_{\chi_1\chi_1\to q\bar q} = \frac{ 
N_fg_\chi'^2 g_q^2 M_{\chi_1}^2
}{
2\pi \left[ (4M_{\chi_1}^2 - M_{Z'}^2)^2 + M_{Z'}^2 \Gamma_{Z'}^2 \right]
} v_{\rm rel}^2 \ .
\end{equation}
Here, the annihilation cross section is $P$-wave suppressed.
A simple way of understanding the $P$-wave nature is from parity. 
The total parity of a fermion-anti-fermion system (applies to two $\chi_1$ particles) is $(-1)^{\ell+1}$, where $\ell$ is the orbital angular momentum between the two particles. 
The axial current (spatial part) has even parity. Therefore we must need $\ell={\rm odd}$ for the annihilation amplitude to be non-vanishing.
It is worth noting that during the thermal freeze out $v_{\rm rel} \simeq \sqrt{6 T_f/M_{\chi_1}}$ and $T_f \sim M_{\chi_1}/25$.  

When the temperature is high enough, $\chi_2$ particles also exist in the universe. As a result, Eq.~(\ref{12int}) permits another annihilation channel, 
the $\chi_1$ and $\chi_2$ coannihilation. The cross section is
\begin{equation}\label{eq:Chi1Chi2Toqqbar}
(\sigma v)_{\chi_1\chi_2\to q\bar q} = \frac{ 
3 N_fg_\chi^2 g_q^2 (M_{\chi_1} + M_{\chi_2})\sqrt{ M_{\chi_1} M_{\chi_2} }
}{
2\pi \left[ ((M_{\chi_1} + M_{\chi_2})^2 - M_{Z'}^2)^2 + M_{Z'}^2 \Gamma_{Z'}^2 \right]
} \ ,
\end{equation}
which is an $S$-wave annihilation. 
For the values of $\delta$ (which controls the $\chi_1$-$\chi_2$ mass splitting) allowed in Fig.~\ref{Validity}, we find $\delta \lesssim T_f$.
Thus, the relative Boltzmann suppression between $\chi_1$ and $\chi_2$ populations is not significant.
In this case, we have $(\sigma v)_{\chi_1\chi_2\to q\bar q} \gg (\sigma v)_{\chi_1\chi_1\to q\bar q}$ because we have chosen $g_\chi\gg g_\chi'$.
At the same time, we also take into account of the $\chi_2\chi_2$ annihilation channels.

There is also another annihilation channel $\chi_1\chi_2 \to Z'Z'$, which  involves one diagonal and one off-diagonal coupling from Eq.~(\ref{12int}),
and is proportional to $g_\chi^2g_\chi'^2$. 
However, we find this cross section is subdominant to $\chi_1\chi_1 \to Z'Z'$ which is an $S$-wave annihilation
contains a $g_\chi^4$ term (Eq.~(\ref{eq:Chi1Chi1ToZ'Z'})), again because $g_\chi\gg g_\chi'$.

\begin{figure}[t]
\centerline{\includegraphics[width=0.6\textwidth]{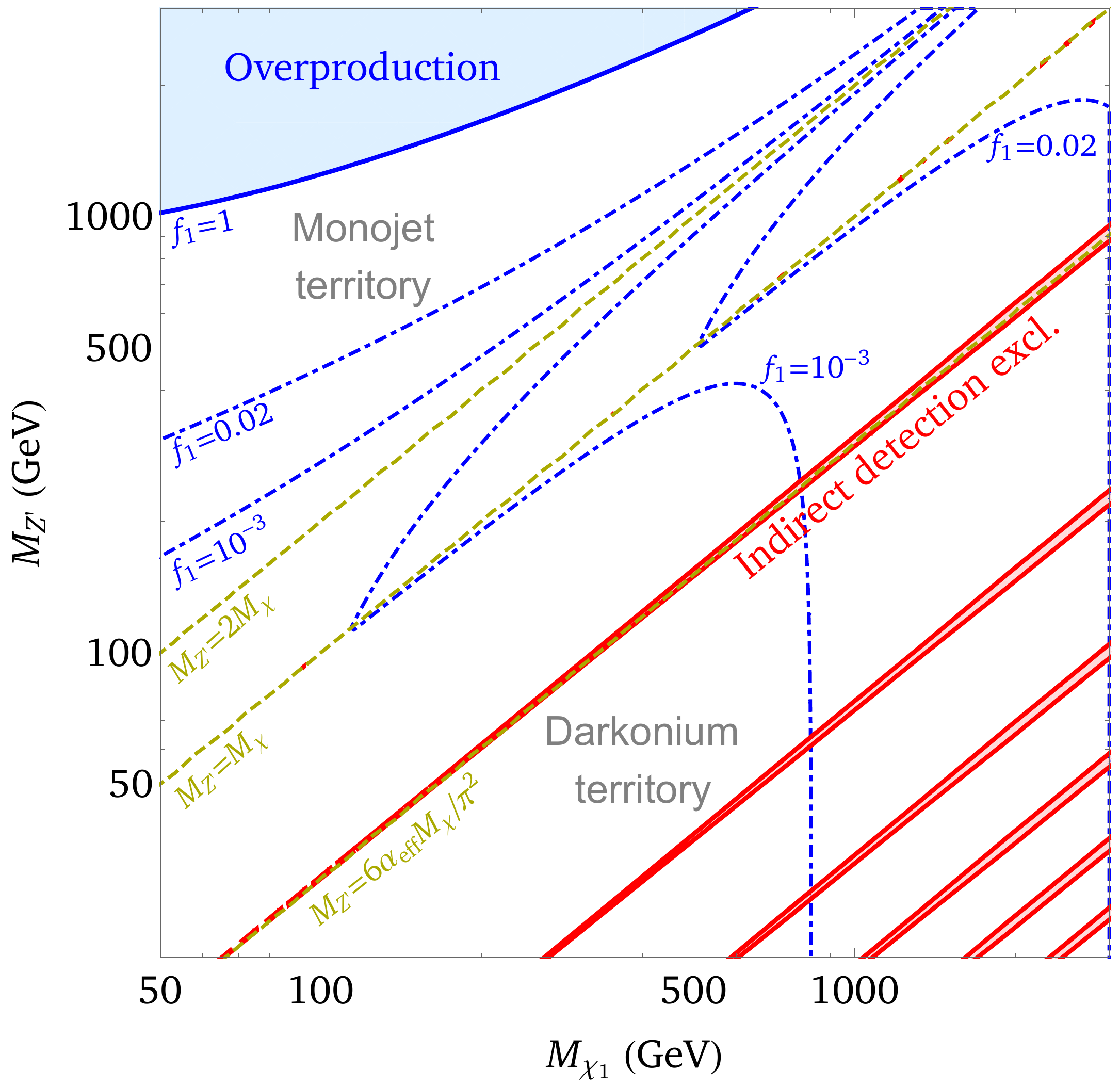}}
\caption{
Cosmological constraints in the same model parameter space as Fig.~\ref{AllConstraints} with the same choice of parameters.
Assuming a thermal history of the dark matter $\chi_1$, it could obtain the correct thermal relic density along the blue solid curve.
The blue shaded region is because of the overproduction of $\chi_1$'s relic density. 
Outside the blue region, $\chi_1$ is underproduced and could only account for a fraction of the total dark matter (shown by blue dot-dashed contours). 
The red spiky regions are ruled out by indirect detection experiments due to the Sommerfeld enhancement,  in spite of the small relic density.
All the white regions in this plot are still alive. 
Here, $f_1$ is the fraction of observed dark matter relic density that is comprised of our dark matter candidate $\chi_1$, defined in Eq.~(\ref{eq:RelicDensity}).
The ``monojet'' and ``darkonium'' territories denote the regions of parameter space where the monojet and darkonium resonance search channels at LHC are most powerful, as discussed in section~\ref{LHC} and shown in Fig.~\ref{AllConstraints}.
}\label{AllConstraintsID}
\end{figure}

In Fig.~\ref{AllConstraintsID}, the blue solid contour shows where $\chi_1$ could obtain the observed dark matter relic abundance~\cite{Ade:2013zuv}, 
by requiring the total annihilation cross section for $\chi_1$ to be equal to $\sigma v_{\rm th} \simeq 3\times 10^{-26}\,{\rm cm^3}/{\rm sec}$. 
We also draw two contours of constant relic density $\chi_1$ in unit of the observed dark matter relic density (labelled by $f_1=0.02$ and $10^{-3}$).
We neglect the non-perturbative Sommerfeld corrections to the cross sections which is usually an order one effect for thermal freeze out.
The shape the blue contours are similar to those found in~\cite{Sirunyan:2017jix}, although in our model we have also kept $g_\chi'$ non-vanishing thus
more annihilation channels have been included.
In the light blue shaded regions, the $\chi_1$ annihilation cross section is smaller than $\sigma_{\rm th}$, thus the dark matter would be overproduced 
in a thermal history. 
Outside the blue shaded regions in Fig.~\ref{AllConstraintsID}, the relic abundance of $\chi_1$ is underproduced.
In this case, its relic density in unit of the observed dark matter relic density is~\footnote{One might resort to non-thermal histories to account for the total observed relic abundance, which is beyond the scope of  current work.}
\begin{equation}\label{eq:RelicDensity}
f_1 \equiv \frac{\Omega_{\chi_1}}{\Omega_{\rm DM}^{obs}} = \frac{\sigma v_{\rm th}}{(\sigma v_{\chi_1})_{\rm tot}} < 1 \ ,
\end{equation}
where $(\sigma v_{\chi_1})_{\rm tot}$ is the sum of the cross sections in Eqs.~(\ref{eq:Chi1Chi1ToZ'Z'})-(\ref{eq:Chi1Chi2Toqqbar}).
This means that $\chi_1$ can only comprise a fraction of the total dark matter in the universe.

%%%%%%%%%%%%%%%%%%%%%%%%%%%%%%%%%%%%%%%%%%%%%%%%%%%%%%%%%%%%%%%%%%%%
\subsection{Indirect detection}\label{ID}
%%%%%%%%%%%%%%%%%%%%%%%%%%%%%%%%%%%%%%%%%%%%%%%%%%%%%%%%%%%%%%%%%%%%

Next, we examine the dark matter indirect detection constraints, assuming $\chi_1$ comprises (a fraction of) the dark matter candidate.
We assume $\chi_2$ do not exist in the universe today.
The indirect signals could arise from $\chi_1\chi_1$ annihilation in the galaxy or the early universe.
We will take into account of the lower limit of $\delta$ derived from direct detection constraints in Eq.~(\ref{eq:MinimalDelta}). 
With a nonzero Majorana mass, one cannot make the assumption that the dark matter relic abundance in the universe is 
asymmetric and argue away the indirect detection constraints~\cite{Buckley:2011ye, Cirelli:2011ac, Tulin:2012re}.

The Born level annihilation cross sections included in this calculation are Eqs.~(\ref{eq:Chi1Chi1ToZ'Z'}) and (\ref{eq:Chi1Chi1Toqqbar}).
For the $P$-wave annihilation, Eq.~(\ref{eq:Chi1Chi1Toqqbar}), it is worth noting that the dark matter halo velocity is a small number, $v_{\rm rel}\sim 10^{-3}$, thus this cross section is highly suppressed.

On top of the Born-level cross sections, we also take into account of the possible Sommerfeld effect in the total annihilation rate. This is especially important when the mass of $Z'$ is smaller compared the de Broglie wavelength of dark matter. We calculate this non-perturbative factor by numerically solving the Schr\"donger equation, following the pioneering works~\cite{Hisano:2004ds, Cirelli:2007xd, ArkaniHamed:2008qn, Pospelov:2008jd, Fox:2008kb, Cassel:2009wt, Feng:2010zp}.

One clarification is necessary with a non-zero Majorana mass $\delta$, where the usual Sommerfeld effect derived for pure Dirac fermion case needs to be modified.
The key picture is that a long-range $Z'$ exchange converts the $\chi_1\chi_1$ initial state into $\chi_2\chi_2$ intermediate state.
Because the typical potential energy is of order $\sim\alpha_\chi^2 M_\chi$, the usual Sommerfeld effect only applies for $\delta\ll\alpha_\chi^2 M_\chi$.
We will assume that this is the case for simplicity. 
If $\delta$ is too large compared to the potential energy, one can no longer cut the ladder diagrams which now becomes genuinely loop suppressed.\footnote{This also corresponds to the green region in Fig.~\ref{Validity}.}
The interplay between $\delta$ and $BE$ in dark matter self interaction was noticed and explored in detail in~\cite{Zhang:2016dck, Blennow:2016gde}.

With the benchmark parameters given in Eq.~(\ref{BenchmarkParameters}), we can evaluate the effective 
cross section for $\chi_1$ annihilation today,
\begin{equation}
(\sigma v)_{\rm eff} = f_1^2 \left[ \mathcal{S} (\sigma v)_{\chi_1\chi_1\to Z'Z'} + \mathcal{S}' (\sigma v)_{\chi_1\chi_1\to q\bar q} \rule{0mm}{4mm}\right] \ ,
\end{equation}
where $\mathcal{S}$ ($\mathcal{S}'$) is the Sommerfeld factor for an $S$- ($P$-) wave annihilation process,
and the factor $f_1^2$ takes into account that $\chi_1$ may only comprise a fraction of the observed dark matter relic density in our model, which is derived based on Eq.~(\ref{eq:RelicDensity}).

An analysis on hidden sector dark matter annihilation has been performed in~\cite{Elor:2015bho} which takes into account that the annihilation into SM particles (quark and antiquarks here) could occur via multiple steps. We adopt the model independent constraints from there.
For the dark matter mass range of interest to this work, the Fermi gamma ray observation from dwarf galaxies~\cite{Ackermann:2015zua} gives the strongest
upper bound on $(\sigma v)_{\rm eff}$. In Fig.~\ref{AllConstraintsID}, the red regions show the parameter space which is ruled out by this indirect measurement.
The spiky feature is mainly due to the Sommerfeld effect.
Clearly, the indirect detection constraint can only exclude very limited regions.
In the region $M_{Z'} < 6 \alpha_{\rm eff} M_{\chi}/\pi^2$ where dark matter bound states exist, we still need the future running of LHC and higher-energy colliders to 
effectively probe this region (see also discussions in section~\ref{sec:DMBSLHC}).% --- This is the main point of this paper.

%%%%%%%%%%%%%%%%%%%%%%%%%%%%%%%%%%%%%%%%%%%%%%%%%%%%%%%%%%%%%%%%%%%%
\section{Conclusion}\label{sec:4}
%%%%%%%%%%%%%%%%%%%%%%%%%%%%%%%%%%%%%%%%%%%%%%%%%%%%%%%%%%%%%%%%%%%%

The nature of dark matter remains mysterious to us after a tremendous amount of effort in searching for them. This strongly suggests going beyond the existing approaches and cast a wide net. 
One important aspect is to broaden the mission of existing experiments. 
In this work, we propose reinterpreting the LHC di-jet (multi-jet) resonance search results to look for darkonium bound states which occur
in dark sector models with a light dark force carrier and a sizable dark coupling with dark matter. 
We focus on a simple model where the dark matter interacts with standard model quarks via the exchange of a vector boson $Z'$,
which is the same benchmark model widely employed by  mono-X searches at the LHC experiments.
In the parameter space where the $Z'$ is weakly coupled to quarks but strongly coupled to the dark matter,
we show that darkonium channel is most useful and highly complementary to mono-X searches. 
Both ought to be included and contrasted in the analysis of future results from LHC and higher energy colliders.

We have also considered the dark matter production in the early universe as well as direct and indirect detection constraints.
We identify the parameter space where these constraints could be weakened, and the reasons behind.
The strong direct detection limits can be evaded by turning on a small Majorana mass for dark matter and split the Dirac fermion into two Majorana
particles. As a consequence, this excludes the possibility of accommodating the asymmetric dark matter scenario thus indirect detection
must be considered. 
To derive the population of our dark matter today, we resort to a thermal history and assume it acquires its relic density
via the freeze out mechanism.
Because the dark gauge coupling of interest to this work is order one (for bound states to exist), for
dark matter below a few TeV, it is underproduced and could only comprise a fraction of the observed dark matter relic density.
This suppresses the indirect detection limits even in the presence of strong Sommerfeld enhancement effects.
We find the above effects occur in a large portion of the model parameter space, where the collider searches 
is the most powerful in probing dark matter in this model.

Although all our findings are based on a very simple model, it is worth emphasizing that the mono-X versus darkonium complementarity as well as
some of the dark matter features derived here are generic and applicable to many extended dark sector models.

%%%%%%%%%%%%%%%%%%%%%%%%%%%%%%%%%%%%%%%%%%%%%%%%%%%%%%%%%%%%%%%%%%%%
\section*{Acknowledgement}
%%%%%%%%%%%%%%%%%%%%%%%%%%%%%%%%%%%%%%%%%%%%%%%%%%%%%%%%%%%%%%%%%%%%

We acknowledge helpful discussions with Kristian Hahn. IL would like to thank Wei Xue for discussions on indirect detection constraints. The work of AK, IL and YZ was supported by US DOE Grant No. DE-SC0010143.  IL is also supported by US DOE grant No. DE-AC02-06CH11357.
YZ acknowledges a generous travel support from the Colegio de F\'isica Fundamental e Interdiciplinaria de las Am\'ericas (COFI) in San Juan, Puerto Rico 
where part of this work has been done.
This manuscript has been authored by Fermi Research Alliance, LLC under Contract No.~DE-AC02-07CH11359 with the U.S. Department of Energy, Office of Science, Office of High Energy Physics. The United States Government retains and the publisher, by accepting the article for publication, acknowledges that the United States Government retains a non-exclusive, paid-up, irrevocable, world-wide license to publish or reproduce the published form of this manuscript, or allow others to do so, for United States Government purposes.

%%%%%%%%%%%%%%%%%%%%%%%%%%%%%%%%%%%%%%%%%%%%%%%%%%%%%%%%%%%%%%%%%%%%%
\appendix
%%%%%%%%%%%%%%%%%%%%%%%%%%%%%%%%%%%%%%%%%%%%%%%%%%%%%%%%%%%%%%%%%%%%%
%
%%%%%%%%%%%%%%%%%%%%%%%%%%%%%%%%%%%%%%%%%%%%%%%%%%%%%%%%%%%%%%%%%%%%%
\section{Origin of a dark matter axial-current coupling to $Z'$}\label{SCPV}
%%%%%%%%%%%%%%%%%%%%%%%%%%%%%%%%%%%%%%%%%%%%%%%%%%%%%%%%%%%%%%%%%%%%%

In this appendix, we present a simple model that could generate an axial current dark matter coupling to the $Z'$.
Under the gauged $U(1)_B$ symmetry, the left- and right-handed components of dark matter $\chi$ have different charges. The charge assignment is

\begin{table}[h]
\centering\begin{tabular}{c|c}
\hline
field  & $U(1)$ charge \\  
\hline
$\chi_L$  &  $q_L$ \\ 
\hline
$\chi_R$  &  $q_R$ \\ 
\hline
$\phi$  &  $q_L-q_R$ \\ 
\hline
\end{tabular}
\end{table}

We assume $q_L\neq q_R$. The Lagrangian that respects $U(1)_B$ takes the following form
\begin{eqnarray}
\mathcal{L} &=& i \bar \chi_L \gamma^\mu (\partial_\mu - i g_B q_L Z'_\mu) \chi_L + i \bar \chi_R \gamma^\mu (\partial_\mu - i g_B q_R Z'_\mu) \chi_R  \nonumber \\
&+& \left[(\partial^\mu - i g_B (q_L-q_R) Z'^\mu) \phi\right]^\dagger \left[(\partial_\mu - i g_B (q_L-q_R) Z'_\mu) \phi\right] + V(\phi) \nonumber \\
&+& y \bar\chi_L \chi_R \phi + {\rm h.c.} \ .
\end{eqnarray}
Because $\chi_L$ and $\chi_R$ have different $U(1)$ charges, we cannot directly write down a mass term, but instead a Yukawa coupling with $\phi$. We assume the scalar potential $V(\phi)$ is such that $\phi$ get a non-zero vacuum expectation value, $\langle \phi \rangle = w/\sqrt2$. This vev breaks the $U(1)$ gauge symmetry giving a mass to $V$ and also give mass to the fermion $\chi$ via the Yukawa coupling.

The particle mass spectrum after the symmetry breaking is
\begin{eqnarray}
M_{Z'} &=& g_B |g_L - g_R| w \ , \nonumber \\
M_\chi &=& y w /\sqrt2 \ .
\end{eqnarray}

The gauge coupling between $\chi$ and $Z'$ can be rewritten as
\begin{eqnarray}
\bar \chi \gamma^\mu \left[ g_B \frac{q_L+q_R}{2} + g_B \frac{q_R-q_L}{2} \gamma_5 \right] \chi Z'_\mu \ .
\end{eqnarray}
Compared to the definition of parameter we have been using, we have
\begin{eqnarray}
g_\chi = g_B \frac{q_L+q_R}{2} \ , \ \ \
g_{\chi}' = g_B \frac{q_R-q_L}{2} \ .
\end{eqnarray}
If the charges $q_L$ and $q_R$ are close to each other, the coupling $g_\chi'$ is suppressed by the difference, so is the mass of the vector boson $M_{Z'} = 2 g_\chi' w$.
In together with the fermion mass, we also find the relation
\begin{eqnarray}
\frac{2g_\chi'}{M_{Z'}} = \frac{y}{\sqrt2 M_\chi} = \frac{1}{w} \ .
\end{eqnarray}

In general the value of the Yukawa coupling $y$ is bounded from above by perturbative unitarity, roughly $y \lesssim \sqrt{4\pi}$. Therefore, we find an upper bound
similar to the one given in Eq.~(\ref{eq:unitarity}),
\begin{eqnarray}\label{upperlimitg'}
g_\chi' \lesssim \frac{\sqrt{\pi} M_{Z'}}{\sqrt2 M_\chi}  \ .
\end{eqnarray}

%%%%%%%%%%%%%%%%%%%%%%%%%%%%%%%%%%%%%%%%%%%%%%%%%%%%%%%%%%%%%%%%%%%%
\bibliography{References}
\bibliographystyle{JHEP}

\end{document}